\documentclass[aps, prx, reprint, longbibliography, nofootinbib,superscriptaddress, floatfix]{revtex4-1}

\usepackage{slashed}
\usepackage{verbatim}
\usepackage[T1]{fontenc}
\usepackage{mathbbol}
\usepackage[dvipsnames]{xcolor}
\usepackage{orcidlink}

\usepackage[normalem]{ulem}
\usepackage[english]{babel}
\usepackage{lipsum}
\usepackage{physics}
\usepackage{dcolumn}
\usepackage{tensor}
\usepackage{comment}
\usepackage{graphicx,color,overpic,mathtools}
\usepackage{amsthm,amsmath,amssymb,mathrsfs}
\usepackage{braket,bm,bbm,setspace}
\usepackage{cancel}
\usepackage{float}
\usepackage{xargs}
\definecolor{myred}{RGB}{179, 27, 27}
\usepackage{hyperref}
\hypersetup{
    colorlinks=true,
    linkcolor=myred, 
    citecolor=myred, 
    urlcolor=myred  
 }
\usepackage{siunitx}
\newcommand{\cgs}{\mathrm{g/cm/s}}

\begin{document}

\title{Axial Oscillations of Viscous Neutron Stars}

\author{Sofía Bussi\`eres\,\orcidlink{0009-0002-8990-4788}}
\affiliation{Center of Gravity, Niels Bohr Institute, Blegdamsvej 17, 2100 Copenhagen, Denmark}
\affiliation{Departament de F\'{\i}sica, Universitat de les Illes Balears, IAC3 - IEEC, Carretera de
Valldemossa km 7.5, E-07122 Palma, Spain}

\author{Jaime Redondo-Yuste\orcidlink{0000-0003-3697-0319}}
\affiliation{Center of Gravity, Niels Bohr Institute, Blegdamsvej 17, 2100 Copenhagen, Denmark}
\affiliation{William H. Miller III Department of Physics \& Astronomy, Johns Hopkins University\\
3400 North Charles Street, Baltimore, MD 21218, USA}

\author{Jos\'e~Javier~Ortega~G\'omez\,\orcidlink{0009-0004-4920-5792}}
\affiliation{Center of Gravity, Niels Bohr Institute, Blegdamsvej 17, 2100 Copenhagen, Denmark}
\affiliation{Instituto de F\'isica Te\'orica (IFT) UAM-CSIC, C/ Nicol\'as Cabrera 13-15, Campus de Cantoblanco UAM, 28049 Madrid, Spain}

\author{Vitor Cardoso\orcidlink{0000-0003-0553-0433}}
\affiliation{Center of Gravity, Niels Bohr Institute, Blegdamsvej 17, 2100 Copenhagen, Denmark}
\affiliation{CENTRA, Departamento de F\'{\i}sica, Instituto Superior T\'ecnico -- IST, Universidade de Lisboa -- UL, Avenida Rovisco Pais 1, 1049-001 Lisboa, Portugal}

\begin{abstract}
The oscillation modes of stars play an important role in observations, and on the understanding of stellar stability properties. The role of viscosity in the oscillation modes of compact stars has been so far understood very loosely only, in absence of a well posed framework. We use recent breakthroughs in the formulation of a causal and stable theory of relativistic hydrodynamics, to study oscillation modes of neutron stars. We characterize the axial spectrum of compact stars and uncover new, viscosity-driven families of modes, without a perfect fluid counterpart. Our results show mode avoidance in some of these families, and a spectrum of long-lived modes, whose role in astrophysical, dynamical processes is yet to be understood. 
\end{abstract}

\maketitle

\section{Introduction}
The Universe is governed by four fundamental interactions: gravity, electromagnetism, and the weak and strong nuclear forces. Often, the details of one or more of these can be safely ignored. For example, black hole mergers are accurately described by gravity alone, while particle collisions at accelerators can ignore gravitational effects. Neutron stars (NSs) are one of the few systems where all four interactions play a major role, shaping their dynamics in crucial ways~\cite{Baym:2017whm}. Our improving ability to study NSs through electromagnetic~\cite{Miller:2019cac, Riley:2019yda, Miller:2021qha, Riley:2021pdl, Bogdanov:2006zd, Bogdanov:2019ixe, Bogdanov:2019qjb} and gravitational-wave (GW) observations~\cite{LIGOScientific:2017vwq, LIGOScientific:2017ync, LIGOScientific:2017zic,LIGOScientific:2018hze, Chatziioannou:2020pqz} therefore singles them out as a uniquely promising laboratory for new physics.

A key goal in NS physics is to determine the composition of cold nuclear matter at supranuclear densities~\cite{Akmal:1998cf, Baym:2017whm}. NICER observations constrain the NS mass-radius relation, and thereby the underlying nuclear equation of state, by measuring X-ray pulse profiles from rotating NSs~\cite{Miller:2019cac, Miller:2021qha, Riley:2019yda, Riley:2021pdl, Bogdanov:2019ixe, Bogdanov:2019qjb}. GW measurements of the tidal deformability provide important complementary constraints~\cite{LIGOScientific:2018hze, LIGOScientific:2018cki}. Looking ahead, third-generation GW detectors such as the Einstein Telescope~\cite{ET:2019dnz, ET:2025xjr} and Cosmic Explorer~\cite{Reitze:2019iox} are expected to resolve the oscillation spectrum of NSs by accessing the post-merger GW emission of binary NS coalescences, opening yet another channel to probe matter under extreme conditions. 

These increasingly precise measurements must be matched by equally accurate theoretical models. To date, most studies do not consider in detail out-of-equilibrium effects such as shear and bulk viscosity (see~\cite{cutler1987effect, cutler1990damping} for earlier work in this direction). While this is a reasonable approximation when NSs remain close to chemical equilibrium, a merger strongly perturbs that equilibrium and can render dissipative processes non-negligible. In particular, if hyperons~\cite{Alford:2017rxf, Alford:2020lla, Most:2021zvc, Most:2022yhe,Ghosh:2023vrx, Ghosh:2025glz} or strange quarks~\cite{Ghosh:2025wfx} are present in the post-merger remnant, weak interaction-mediated reactions (e.g., modified Urca processes) can dramatically enhance bulk viscosity~\cite{Alford:2018lhf}. High-precision GW asteroseismology must therefore account for viscous effects in NS oscillations; conversely, future GW observations may allow us to infer the underlying viscous transport coefficients of dense matter. Viscosity can also leave imprints in the dynamical tidal deformability during the late inspiral~\cite{HegadeKR:2024agt, HegadeKR:2024slr, Ripley:2023lsq, Ripley:2023qxo, HegadeKR:2026iou}.

Recent efforts laid the groundwork to study NS oscillations including both relativistic effects and dissipation~\cite{Redondo-Yuste:2024vdb,Diaz-Guerra:2024gff}. In particular, the scattering of GWs off non-rotating~\cite{Boyanov:2024jge} and slowly rotating stars~\cite{Redondo-Yuste:2025ktt} has been investigated, demonstrating the absorption and superradiant amplification of GWs by viscous stars. The role of viscosity in radial oscillations of NSs has been recently analyzed in Refs.~\cite{Mendes:2025oib, Caballero_2025, Shum:2025jnl,Keeble:2026bzo}. In parallel, major progress has been made in implementing dissipative effects in numerical simulations of relativistic hydrodynamics~\cite{Pandya:2022pif, Pandya:2022sff, Bantilan:2022ech, Keeble:2025bkc,Clarisse:2025lli,Bea:2025eov, Chomali-Castro:2026qww}, and specifically neutron star mergers~\cite{Chabanov:2023abq, Chabanov:2023blf, Chabanov:2024yqv}. Building on these developments, we provide the first systematic study of how viscosity affects a family of non-radial oscillation modes of NSs. As a first step, we focus on axial perturbations: demonstrating how shear viscosity damps and modifies the spacetime or $w$-modes~\cite{Kokkotas:1999bd}, and how a new family of modes, which we dub $\eta$-modes, where the restoring force is provided by shear viscosity, emerges. We envisage this as a step towards accurately modeling the impact of viscosity on the postmerger emission of neutron star mergers. 

Our work builds upon Refs.~\cite{Redondo-Yuste:2024vdb, Boyanov:2024jge} to study the coupled perturbations of the spacetime metric with a dissipative fluid. Viscosity is modeled within a causal and stable theory of relativistic hydrodynamics, in the framework introduced by Bemfica, Disconzi, Noronha~\cite{Bemfica:2017wps,Bemfica_2018, Bemfica_2019, Bemfica:2019knx, Bemfica:2020zjp}, and Kovtun~\cite{Kovtun_2019}, dubbed BDNK Hydrodynamics. The main advantage of this approach is that the resulting theory only contains first gradients of the thermodynamic variables, which in turn simplifies the treatment of the perturbed Einstein-hydro equations, compared to the more complicated second-order theories of the Israel-Stewart class~\cite{Israel:1979wp}. 

We describe the framework of dissipative, relativistic hydrodynamics we consider in Section~\ref{sec:framework}. Next, we explain in section~\ref{sec:methods} how do we obtain the quasinormal mode frequencies. Our results are presented in section~\ref{sec:Results}. All throughout our work we use geometric units $G=c=1$ unless otherwise specified, lower-case latin letters denote spacetime indices, $a,b=0,\dots,3$, and we use the mostly plus metric signature $(-+++)$.

\section{Framework\label{sec:framework}}
We study perturbations of spherically symmetric stars. For simplicity, we focus on the axial or odd-parity sector. The star is sourced by a dissipative fluid, modeled within a causal and well-posed formulation of relativistic Navier-Stokes hydrodynamics. 
\subsection{Relativistic Navier-Stokes}
For most purposes regarding NS dynamics, it is a good approximation to consider that the stars are cold, and diffusion is negligible (vanishing chemical potential). In that case, the matter is barotropic, and the equation of state is of the form $p=p(\rho)$, with $p,\rho$ denoting the pressure and energy density, respectively. Hydrodynamics is described by an effective stress energy tensor $T_{ab}[\rho, u^a]$ which depends on the energy density $\rho$ and the fluid $4$-velocity $u^a$ (normalized such that $u^au_a=-1$). 

Generically, one may think of hydrodynamics as an effective field theory~\cite{Kovtun:2012rj}. Out of equilibrium effects can be captured by including order-by-order terms that depend on higher-order derivatives of the fundamental variables. In that sense, the stress energy tensor can be written as $T_{ab} = T_{ab}^{(0)} + T^{(1)}_{ab} + \dots$, where $T_{ab}^{(0)}$ is the stress-energy tensor of a perfect fluid, $T^{(1)}_{ab}$ depends on first-order derivatives of $\rho$ and $u^a$, and so forth. The perfect fluid stress energy tensor is uniquely fixed by the symmetry of the stress energy tensor
\begin{equation}\label{eq:perfect_fluid}
    T_{ab}^{(0)} = \rho u_au_b + p \Delta_{ab} \, ,
\end{equation}
where $\Delta_{ab}=g_{ab}+u_au_b$.
There is no unique prescription for the first-order correction. In general, it takes the form 
\begin{equation}\label{eq:T1}
    T_{ab}^{(1)} = \mathcal{A}u_a u_b + \Pi \Delta_{ab} + 2u_{(a}\mathcal{Q}_{b)} - 2\eta\sigma_{ab} \, , 
\end{equation}
where $\eta$ is the shear viscosity, $\sigma_{ab}$ is the shear tensor given by 
\begin{equation}
    \sigma^{ab}=\frac{1}{2}\Delta^{ac}\Delta^{bd}\left(\nabla_{c}u_{d}+\nabla_{d}u_{c}-\frac{2}{3}\Delta_{cd}\nabla_{e}u^{e}\right) \, , 
\end{equation}
and $\mathcal{A},\Pi,\mathcal{Q}_a$ are functions of $\nabla_a\rho, \nabla_b u_a$. The only requirement is that the heat flux is transverse, $\mathcal{Q}_a u^a = 0$. The simplest prescription one can imagine, simply promoting to the relativistic case the familiar Navier-Stokes hydrodynamics, was proposed by Landau and Lifshitz~\cite{Landau:1987}, and consists on 
\begin{equation}\label{eq:Landau}
    \mathcal{A}^{\rm Landau} = \mathcal{Q}_a^{\rm Landau} = 0 \, , \qquad \Pi^{\rm Landau} = -\zeta \nabla_a u^a \,, 
\end{equation}
where $\zeta$ is the bulk viscosity. This theory, however, has a number of undesirable features, including acausal propagation of perturbations and instability of global equilibrium solutions~\cite{Hiscock:1983, Hiscock:1985}. For this reason, first-order theories were long abandoned and replaced by theories that include second-order corrections (a term of the form $T_{ab}^{(2)}$), such as Israel-Stewart hydrodynamics~\cite{Israel:1979wp}. 

An alternative viewpoint, proposed by BDNK~\cite{Bemfica:2020zjp, Kovtun_2019}, is to consider transformations of the hydrodynamic frame $(\rho,u^a) \to (\rho + \mathcal{O}(\partial), u^a + \mathcal{O}(\partial))$, such that the resulting effective stress energy tensor leads to well-posed equations of motion. Indeed, consider the following frame transformation acting on the Landau frame~\eqref{eq:Landau}
\begin{equation}
    \begin{aligned}
        \rho \to& \rho + \theta \Bigl[u^a\nabla_a \rho + (\rho+p)\nabla_au^a\Bigr] \, , \\
        u^a \to& u^a + \frac{\tau-\theta}{2(\rho+p)}\Bigl[u^a\nabla_a \rho + (\rho+p)\nabla_au^a\Bigr]u^a \\
        &+\tau \Bigl[u^b\nabla_b u^a + \frac{c_s^2}{\rho+p}\Delta^{ab}\nabla_b \rho\Bigr] \, , 
    \end{aligned}
\end{equation}
with $\tau,\theta$ two new transport coefficients with units of length. The new stress-energy tensor, neglecting higher-order in gradient terms, is $T_{ab} = T_{ab}^{(0)}+T_{ab}^{(1)} + \mathcal{O}(\partial^2)$, with $T_{ab}^{(0)}$ given by~\eqref{eq:perfect_fluid} , and $T_{ab}^{(1)}$ given by~\eqref{eq:T1} with constitutive relations
\begin{equation}\label{eq:constitutive}
    \begin{aligned}
        \mathcal{A}=& \tau \Bigl[u^a\nabla_a\rho + (\rho+p)\nabla_au^a\Bigr]  \, , \\
        \Pi =& c_s^2 \theta \Bigl[u^a\nabla_a\rho + (\rho+p)\nabla_au^a\Bigr] -\zeta \nabla_au^a \, , \\
        \mathcal{Q}_a =& \tau \Bigl[(\rho+p)u^b\nabla_b u^a + c_s^2\Delta^{ab}\nabla_b \rho\Bigr] \, ,
    \end{aligned}
\end{equation}
where $c_s^2 = dp/d\rho$ is the sound speed. This is simply a restricted case of the constitutive relations of BDNK hydrodynamics, written in~\cite{Bemfica_2018, Bemfica:2020zjp, Kovtun_2019}.

This theory is causal and stable as long as $(\theta,\tau)$ satisfy certain constraints. It is useful to introduce dimensionless parameters in the problem. In order to test different parametrizations of the viscosity -- these, ultimately, should be informed by the underlying microscopic theory -- we test two different parametrizations, which we label ``A'' and ``B'':
\begin{subequations}\label{eq:viscous_parametrizations}
\begin{align}
\text{A:}\quad 
\eta &= \hat{\eta} (\rho+p) L_0 c_s^2, 
& \theta &= L_0 \hat{\eta}, 
& \tau &= \hat{\tau} L_0 \hat{\eta}, 
\label{eq:par_a} \\[0.5em]
\text{B:}\quad 
\eta &= \hat{\eta} p L_0, 
& \theta &= L_0 \frac{p}{\rho}, 
& \tau &= \hat{\tau} L_0 \frac{p}{\rho}.
\label{eq:par_b}
\end{align}
\end{subequations}
where we introduced a lengthscale $L_0$, and we are setting $\zeta=0$ in what follows. As discussed in~\cite{Redondo-Yuste:2024vdb}, the bulk viscosity does not affect the perturbations in the axial sector, so we can safely ignore it. The transport coefficients are parametrized by two dimensionless constants $\hat{\eta}, \hat{\tau}$. This is a causal frame if the following conditions are met:
\begin{equation}
    \begin{aligned}
        \text{A:}& \quad \hat{\eta}\geq 0 \,  ,\quad \hat{\tau} > 0 \, , \quad 0 \leq c^2_s \leq \frac{\hat{\tau}}{2+\hat{\tau}} \, , \\
        \text{B:}& \quad 0\leq\hat{\eta}\leq \frac{3}{4} \,  ,\quad \hat{\tau} > \max\Bigl(\hat{\eta},\frac{2}{1-c_s^2}\Bigr)  .
    \end{aligned}
\end{equation}
Clearly, parametrization A has the advantage that the shear viscosity is not bounded from above. Parametrization B had been previously used in Ref.~\cite{Boyanov:2024jge}, allowing us a direct comparison as well as to test for the independence of our results on the parametrization of the transport coefficients. In both cases, we note that $\hat{\tau}$ needs to be large enough to regulate the maximum sound speed inside the star (typically achieved at the center). The frames that we consider in this work, including the maximum sound speed that they can resolve while preserving causality, are summarized in Table~\ref{tab:frames}.
\begin{table}[t]
\begin{ruledtabular}
\begin{tabular}{lccc}
Frame & Parametrization & $\hat{\tau}$ & $\max c_s^2$ \\
\hline
A1 & A $\sim$ Eq.~\eqref{eq:par_a} & 10  &  $0.83$\\
A2 & A $\sim$ Eq.~\eqref{eq:par_a} & 20 &  $0.91$\\
B1 & B $\sim$ Eq.~\eqref{eq:par_b} & 10 &  $0.8$\\
B2 & B $\sim$ Eq.~\eqref{eq:par_b} & 20 &  $0.9$\\
\end{tabular}
\caption{Summary of the hydrodynamic frames considered in this work. The right column indicates the maximum sound speed for which causality is guaranteed. }
\label{tab:frames}
\end{ruledtabular}
\end{table}

The stress-energy tensor~\eqref{eq:perfect_fluid}-\eqref{eq:T1} (with constitutive relations given by~\eqref{eq:constitutive}) genuinely describes a dissipative fluid. The entropy density current $T S^a = p u^a - T_{ab}u_b$ (with $T$ the temperature) satisfies 
\begin{equation}\label{eq:second_law}
    \nabla_a S^a \geq \frac{2\eta\sigma_{ab}\sigma^{ab}}{T} + \mathcal{O}(\partial^3)  \, . 
\end{equation}
The first term is clearly non-negative as long as $\eta\geq 0$. Hence, the second law of thermodynamics is satisfied as long as the higher-order terms remain small, i.e., as long as we remain within the regime of validity of first-order hydrodynamics. 

In the following, we will consider large values of the shear viscosity, of the order of $\eta_c \sim 10^{29-32} \cgs$, corresponding approximately to $\hat{\eta}\sim 10^{-3}-1$ in the above parametrizations. This is unrealistic for a neutron star -- shear viscosity arising from the scattering of quarks or electrons can be as large as $\eta \sim 10^{20}\cgs$ at temperatures of $T\sim 10^{7}\mathrm{K}$. We do this for two reasons: (i) smaller shear viscosities do not lead to any observable effect, and (ii) the bulk viscosity of a neutron star featuring strange matter in the core can be as large as $\zeta \sim 10^{30}\cgs$~\cite{Ghosh:2025glz,Ghosh:2025wfx}, so our results can be seen as an estimate of the effect of a large transport coefficient in the stellar oscillation modes. Finally, when discussing ultracompact objects, notice that their shear viscosity must satisfy the Kovtun-Son-Starinets (KSS) bound $\eta \geq s/(4\pi)$, with $s$ the (volume) entropy density~\cite{Kovtun:2003wp, Kovtun:2004de}. A genuine black hole mimicker ought to have a similar entropy density as a black hole, $s \sim s_{\rm BH}$, so we can estimate its shear viscosity to be $\eta \gtrsim 1/(16\pi\ell_{\rm eff})$, with $\ell_{\rm eff} \gtrsim M$ an effective lengthscale introduced to estimate the \emph{volumetric} entropy density. The KSS bound then gives an estimate of the shear viscosity of a compact object of 
\begin{equation}
    \eta \gtrsim 5.4 \times 10^{31}\cgs \Bigl(\frac{M_\odot}{M}\Bigr) \, .
\end{equation}
%
\subsection{Equilibrium Solutions}
The metric of equilibrium solutions can be written, in area gauge, as 
\begin{equation}\label{eq:background_metric}
    ds^2 = -e^{\nu}dt^2 + e^{\lambda}dr^2 + r^2d\Omega^2 \, , 
\end{equation}
where $\nu,\lambda$ are purely radial functions. The fluid energy density $\rho$ is also static, and the $4$-velocity is simply $u^\mu = e^{-\nu/2}(\partial_t)^\mu $. Einstein equations reduce to the Tolman-Oppenheimer-Volkoff (TOV) equations~\cite{Tolman_RTC}:
\begin{equation}\label{eq:tov}
    \begin{aligned}
        m' =& 4\pi r^2\rho \, , \\
        \nu' =& \frac{2m+8\pi r^3 p}{r(r-2m)} \, , \\
        p' =& -\frac{(\rho+p)(m+4\pi r^3 p)}{r(r-2m)} \, ,
    \end{aligned}
\end{equation}
where $e^{-\lambda}=1-2m/r$ and primes denote radial derivatives. 

We integrate these equations starting from some central value of the energy density $\rho(r=0)=\rho_c$, up to the surface radius $R$, defined as the point where $p(r=R)=0$. At that point, we match with a Schwarzschild solution with mass $M = m(R)$. 

We consider two different polytropic equations of state, of the form
\begin{equation}\label{eq:EoS}
    p = \kappa \rho^{1+1/n} \, , 
\end{equation}
as well as constant density stars in Section~\ref{ssec:ucos}. In particular, we will choose $\kappa=100\,\mathrm{km}^2, n=1$, and $\kappa=700\,\mathrm{km}^{2.5}, n=0.8$.  Being analytic, these equations of state allow for a simple numerical implementation. Unless otherwise specified, we consider a central density of $\rho_c = 3\times 10^{15}\mathrm{g/cm^3}$. This leads to a star with $M=1.27M_\odot$ and $R=8.86\mathrm{km}$ for the first equation of state, and $M=1.54M_\odot$, $R=8.78\mathrm{km}$ for the latter. However, the method discussed in this work is readily applicable to generic equations of state.  

\subsection{Axial Modes}
Now we consider the dynamics of the fluid-gravity system under small deviations with respect to equilibrium. We assume that the metric and stress energy tensor admit a formal perturbative expansion of the form
\begin{equation}
    g_{\mu\nu}= \bar{g}_{\mu\nu} + \varepsilon h_{\mu\nu} \, , \qquad T_{\mu\nu} = \bar{T}_{\mu\nu} + \varepsilon \delta T_{\mu\nu} \, , 
\end{equation}
with $\varepsilon \ll 1$, where $\bar{g},\bar{T}$ denote the background metric~\eqref{eq:background_metric} and stress-energy tensor, obtained as solutions of the TOV equations~\eqref{eq:tov}. As the background is spherically symmetric, we can decompose the perturbed tensor in tensor spherical harmonics with well-defined parity. Even and odd parity perturbations decouple. In this work, we focus only on odd parity (axial) perturbations, first studied in Ref.~\cite{1967ApJ...149..591T}. Odd metric perturbations in the Regge-Wheeler gauge reduce to 
\begin{equation}
    h_{aA} = \sum_{\ell,m} h_a(r)e^{-i\omega t} X_A(\theta,\phi) \, , \quad a=t,r,\, \,  A=\theta,\phi \, .
\end{equation}
The only axial perturbations to the stress energy tensor are those generated by perturbations of the fluid velocity, which decompose as 
\begin{equation}
    \delta u^A  = e^{-\nu/2}e^{-i\omega t} \frac{Z(r)}{r^2}X^A(\theta,\phi)  \, , \qquad A=\theta,\phi \, .
\end{equation}
The linearized Einstein equations lead to a coupled system of equations for $(h_t,h_r,Z)$. Eliminating $h_t$, and defining $\psi = e^{(\nu-\lambda)/2}h_r/r$, the problem reduces to two coupled ODEs~\cite{Redondo-Yuste:2024vdb}
\begin{align}
    f\Bigl[d(f \psi')\Bigr]' + (\omega^2-V)\psi =& -16\pi e^{\nu/2} i\omega\eta \psi + C_1 Z\, , \label{eq:gw_eq}\\
    f\Bigl[d(f Z')\Bigr]' + (c_\eta^{-2} \omega^2-U)Z =& C_2 Z' + C_3 Z + C_4 \psi' + C_5 \psi \label{eq:fluid_eq}\, ,
\end{align}
where we have introduced for convenience $f^2=e^{\nu-\lambda}$, $V$ is the usual Regge-Wheeler potential for a perfect fluid,
\begin{equation}
    V = e^{\nu}\Bigl[\frac{\ell(\ell+1)}{r^2}-\frac{6m}{r^3} + 4\pi (\rho-p)\Bigr] \, ,\label{RW_potential_star}
\end{equation}
the viscous propagation speed (second-sound) is
\begin{equation}
    c^2_\eta = \frac{\eta}{\tau(p+\rho)} \, ,
\end{equation}
and the functions $U, C_1,\dots,C_5$ are written below:
\begin{widetext}
    \begin{equation}
        \begin{aligned}
            U =& e^{\nu}\Bigl[\frac{\ell(\ell+1)}{r^2}-\frac{2m}{r^3}+8\pi (2p+\rho)\Bigr] \, , \quad
            C_1 = \frac{8\pi e^{\nu-\lambda/2}}{r^2}\Bigl[2r \eta' +\Bigl(e^\lambda(1+8\pi r^2 p)-1\Bigr)\eta\Bigr] \, , \\
            C_2 =& \frac{f^2}{2r}\Bigl[e^\lambda(1+8\pi r^2 p)-1 - \frac{2r\eta'}{\eta}\Bigr] \, , \quad
            C_3 = -i\omega (p+\rho)e^{\nu/2}\Bigl(\frac{1}{\eta}+16\pi\tau\Bigr)+\frac{2f^2\eta'}{r\eta} \, , \\
            C_4 =& r f \Bigl[i\omega + \frac{p+\rho}{\eta}\Bigl(e^{\nu/2}-i\omega\tau\Bigr)\Bigr] \, , \quad
            C_5 = f\Bigl[\frac{(p+\rho)e^{\nu/2}}{\eta} -\frac{i\omega}{2}\Bigl(-7+e^\lambda(1+8\pi r^2 p)\Bigr)+\frac{i\omega}{\eta}\Bigl(r \eta' - (p+\rho)\tau \Bigr)\Bigr] \, .
        \end{aligned}
    \end{equation}
\end{widetext}
Notice that, for parametrization A, the viscous propagation speed is $c^2_\eta = c^2_s/\hat{\tau}$. This means that this second sound speed satisfies 
\begin{equation}
    c^2_\eta \leq c^2_s \iff \max c^2_s \geq \frac{1}{3} \, .
\end{equation}
For all equations of state considered here this will be satisfied. Since the transport coefficients do not explicitly depend on the sound speed in parametrization B, we cannot make such a direct comparison in that case. 

In the perfect fluid limit, multiplying eq.~\eqref{eq:fluid_eq} by $\eta$, and then setting $\eta,\tau=0$, we find that the left hand side vanishes, and the right hand side becomes simply 
\begin{equation}
    i\omega Z= f(r\psi' +\psi) \, .
\end{equation}
Hence, for a perfect fluid, $Z$ can be eliminated in terms of $\psi$, and eq.~\eqref{eq:gw_eq} becomes a decoupled wave equation for $\psi$ (see~\cite{Redondo-Yuste:2024vdb} for further details). 

In the exterior of the star, the only dynamics is given by propagating axial GWs. These are governed by the vacuum Regge-Wheeler equation. Physical solutions are required to be regular at the origin, which enforces $\psi,Z \sim r^{\ell+1}$. Moreover, GWs must be outgoing as $r\to\infty$, $\psi \sim e^{i\omega r_*}$, with $r_* =r +2M\log(r/2M-1)$ the tortoise coordinate outside the star. Finally, we need to match the solution across the surface of the star. Israel junction conditions are trivially satisfied as long as the shear viscosity vanishes at the star's surface $\eta \xrightarrow{r\to R} 0$. Both parametrizations~\eqref{eq:viscous_parametrizations} naturally enforce that $\eta\to 0$ as long as the energy density vanishes at the surface of the star. However, the vanishing of the viscosity at the surface introduces terms that diverge at the surface in eq~\eqref{eq:fluid_eq}. Regular solutions must be such that these divergent terms cancel out, which lead to a regularity condition at the surface. This condition depends on the parametrization chosen, and is given by 
\begin{equation}\label{eq:surface}
    B^{(I)}_1 Z(R) + B^{(I)}_2 Z'(R) + B^{(I)}_3\psi(R) + B^{(I)}_4 \psi'(R) = 0  \, ,  
\end{equation}
with $I=A,B$ denoting the two possible parametrizations. For parametrization A, we find 
\begin{equation}
    \begin{aligned}
        B^{(A)}_1 =& \frac{-4 \mathcal{C}^2 \hat{\eta }+2 \mathcal{C} \hat{\eta }+R \omega  \left(\hat{\eta } \hat{\tau } R \omega +i \sqrt{1-2 \mathcal{C}}\right)}{\hat{\eta }R^2 (1-2 \mathcal{C}){}^2} \, , \\
        B^{(A)}_2 =& \frac{\mathcal{C}}{2R(1-2\mathcal{C})} \, , \\
        B^{(A)}_3 =& \frac{i \left(i (1-2 \mathcal{C})^{3/2}-2 \mathcal{C} \hat{\eta } \hat{\tau } R \omega +\hat{\eta } \hat{\tau } R \omega +\mathcal{C} \hat{\eta } R \omega \right)}{(1-2 \mathcal{C})^2 \hat{\eta } R} \, , \\
        B^{(A)}_4 =& -\frac{1}{\sqrt{1-2 \mathcal{C}} \hat{\eta }}+\frac{i \hat{\tau } R \omega }{1-2 \mathcal{C}}\, , 
    \end{aligned}
\end{equation}
with $\mathcal{C}=M/R$ the star's compactness. For parametrization B, instead, we find
\begin{equation}
    \begin{aligned}
        B^{(B)}_1 =& -(1 - 2 \mathcal{C})^{1/2} i \omega + \frac{\mathcal{C}(5 \mathcal{C}-2)}{R ^2} \hat{\eta}, \\
        B^{(B)}_2 =& -\mathcal{C} \hat{\eta} (1 - 2 \mathcal{C}), \\
        B^{(B)}_3 =& - \frac{M}{(1 - 2 \mathcal{C})^{1/2}} \hat{\eta} i \omega + 1 - 2 \mathcal{C}, \\
        B^{(B)}_4 =& R (1 - 2 \mathcal{C})
    \end{aligned}
\end{equation}

We emphasize that this is \emph{not} a boundary condition: the only boundary condition is that the GWs are outgoing at $r\to \infty$. This condition simply ensures the regularity of the solutions to the coupled equations~\eqref{eq:gw_eq}-\eqref{eq:fluid_eq}.   
\section{Methods\label{sec:methods}}
Quasinormal modes are the discrete frequencies $\omega$ for which eqs.~\eqref{eq:gw_eq}-\eqref{eq:fluid_eq} admit regular solutions which are outgoing at infinity. Our strategy to find these frequencies is, in essence, to solve the wave equations satisfying both regularity conditions (at the inside and at the surface), and the outgoing boundary condition. Only for the quasinormal modes frequencies will both solutions match smoothly, allowing us to formulate the problem as a root-finding problem for the frequency $\omega$. 

In detail, we find two solutions: $\psi_{\rm in}, \psi_{\rm up}$, where $\psi_{\rm in}$ is regular at the origin and at the surface of the star, and $\psi_{\rm up}$ is an outgoing solution to the equation in the exterior of the star. Quasinormal modes correspond to the values of $\omega$ such that the two solutions are not linearly independent. We can quantify this by requiring the vanishing of the Wronskian
\begin{equation}\label{eq:wronskian}
    \Delta(\omega) \equiv \psi_{\rm in}(a)\psi_{\rm up}'(a) - \psi_{\rm up}(a) \psi_{\rm in}'(a) = 0  \, ,
\end{equation}
at some matching radius $a\geq R$. 
\subsection{In Solution}
In order to find the in-solution $(\psi_{\rm in}, Z_{\rm in})$ we first integrate the coupled system~\eqref{eq:gw_eq}-\eqref{eq:fluid_eq} from $r=r_{\rm min} \sim 0$, with boundary conditions that explicitly enforce regularity. We consider two independent solutions, characterized by the boundary conditions
\begin{equation}
    \begin{aligned}
        \psi_{\rm in}^{(1)}(r_{\rm min}) =& r_{\rm min}^{\ell+1} + \mathcal{O}(r_{\rm min}^{\ell+2}) \, , \qquad Z^{(1)}_{\rm in}(r_{\rm min}) = 0 \, , \\ 
        \psi_{\rm in}^{(2)}(r_{\rm min}) =& 0 \, , \qquad Z^{(2)}_{\rm in}(r_{\rm min}) = r_{\rm min}^{\ell+1} + \mathcal{O}(r_{\rm min}^{\ell+2})\, .
    \end{aligned}
\end{equation}
where the higher order terms can be found analytically by expanding the equations near the origin. Regular solutions at the surface can always be written as a linear combination of these two independent solutions, 
\begin{equation}
    \psi_{\rm in} = \psi_{\rm in}^{(1)} + K \psi_{\rm in}^{(2)} \, , \qquad Z_{\rm in} = Z_{\rm in}^{(1)} + K Z_{\rm in}^{(2)} \, .
\end{equation}
Inserting the above ansatz into the regularity condition~\eqref{eq:surface} leads to 
\begin{equation}
    K = -\frac{B_1 Z_{\rm in}^{(1)} + B_2 (Z_{\rm in}^{(1)})' + B_3 \psi_{\rm in}^{(1)} + B_4 (\psi_{\rm in}^{(1)})'}{B_1 Z_{\rm in}^{(2)} + B_2 (Z_{\rm in}^{(2)})' + B_3 \psi_{\rm in}^{(2)} + B_4 (\psi_{\rm in}^{(2)})'}\Biggr|_{r=R} \, .
\end{equation}
This uniquely fixes (up to proportionality constants) the interior solution $\psi_{\rm in}$. We now continue the integration outwards, by solving the exterior Regge-Wheeler equation
\begin{equation}\label{eq:exterior}
    f\Bigl[d(f \psi')\Bigr]' + (\omega^2-V)\psi = 0 \, , \qquad f=1-\frac{2M}{r} \, , 
\end{equation}
where $V$ is the Regge-Wheeler potential \eqref{RW_potential_star} evaluated in vacuum, i.e. 
\begin{equation}
    V =f\Bigl[\frac{\ell(\ell+1)}{r^2}-\frac{6M}{r^3}\Bigr] \, .
\end{equation}
The boundary condition at $r=R$ is given by requiring continuity with the previous solution. We extend this integration up to the matching radius $r=a$. 

\subsection{Up Solution}
We follow~\cite{Benhar:1998au} to find the solution at a radius close to the surface of the star in terms of a continued fraction expansion. Notice that as opposed to Ref.~\cite{Benhar:1998au}, the imaginary part of the frequency corresponding to stable modes is taken to be negative, i.e., $\Psi(t,r) = e^{-i\omega t}\psi$. 

Let $v=1-a/r$, and let
\begin{equation}\label{eq:field_redefinition}
    \psi =\chi(r)\phi(v) \, , \quad \chi(r) = (r-2M)^{2i\omega M}e^{i\omega r} \, .
\end{equation}
A straightforward calculation shows that $\phi$ satisfies the following equation
\begin{widetext}
    \begin{equation}
    \begin{aligned}
        &(c_0+c_1v+c_2v^2+c_3v^3)\phi_{,vv} + (d_0+d_1v+d_2v^2)\phi_{,v} + (e_0+e_1v)\phi = 0 \, , \\ 
        c_0 =& 1-\frac{2M}{a} \, , \quad c_1  = \frac{6M}{a}-2  \, , \quad c_2 = 1-\frac{6M}{a} \, , \quad c_3 = \frac{2M}{a} \, , \\
        d_0 =& \frac{6M}{a}-2(1-ia\omega) \, , \quad d_1 = 2-\frac{12M}{a} \, , \quad d_2 = \frac{6M}{a} \, , \quad
        e_0 = \frac{6M}{a} -\ell(\ell+1) \, , \quad e_1 = -\frac{6M}{a} \, .
    \end{aligned}
\end{equation}
\end{widetext}
We can write a power-series solution for $\phi$ in terms of a four-term recurrence relation
\begin{equation}
    \begin{aligned}
        \phi =& \sum_{n=0}^\infty \phi_n v^n \, , \\ \alpha_n \phi_{n+1}&+\beta_n\phi_n + \gamma_n\phi_{n-1}+\delta_n\phi_{n-2}=0 \, , \quad n\geq 2 \, ,
    \end{aligned}
\end{equation}
with 
\begin{equation}
    \begin{aligned}
        \alpha_n =& n(n+1)c_0 \, , \\
        \beta_n =& n(n-1)c_1 + nd_0 \, , \\
        \gamma_n =& (n-1)(n-2)c_2 + (n-1)d_1 + e_0 \,,  \\
        \delta_n =& (n-2)(n-3)c_3 + (n-2)d_2 + e_1 \, .
    \end{aligned}
\end{equation}
This relation fixes uniquely all higher order coefficients in terms of $(\phi_0,\phi_1)$. Indeed, evaluating $\psi$ at $r=a$, corresponding to $v=0$, we find 
\begin{equation}
    \phi_0 = \frac{\psi(a)}{\chi(a)} \, , \quad \phi_1 = \frac{a}{\chi(a)}\Bigl(\psi'(a) + \frac{i\omega a}{a-2M}\psi(a)\Bigr) \, .
\end{equation}
This allows us now to invert the recurrence relation in terms of a continued fraction, following~\cite{Leaver:1985ax, Leaver:1986vnb}. A key step in order to do that is to reduce the four-term recurrence relation to a three-term recurrence relation~\cite{Leaver:1990zz}. This is easily done by defining 
\begin{equation}
    \hat{\alpha}_n = \alpha_n \, , \quad \hat{\beta}_n = \beta_n - \frac{\delta_n\hat{\alpha}_{n-1}}{\hat{\gamma}_{n-1}} \, , \quad \hat{\gamma}_n = \gamma_n - \frac{\delta_n\hat{\beta}_{n-1}}{\hat{\gamma}_{n-1}} \, , 
\end{equation}
leading to the three-term recurrence 
\begin{equation}
    \hat{\alpha}_n\phi_{n+1}+\hat{\beta}_n\phi_n+\hat{\gamma}_{n}\phi_{n-1} = 0 \, .
\end{equation}
Key results for three term recursion relations are summarized in Ref.~\cite{Gautschi:1967}. For our purposes, suffices to know that they admit a \emph{minimal} solution which is absolutely and uniformly convergent outside the star as long as $4M < a < 2R < 2a$. If the recurrence relation admits a minimal solution, then the following continued fraction holds
\begin{equation}
    \frac{\phi_1}{\phi_0} = - \frac{\hat{\gamma}_1}{\hat{\beta}_1-}\frac{\hat{\alpha}_1\hat{\gamma}_2}{\hat{\beta}_2-}\frac{\hat{\alpha}_2\hat{\gamma}_3}{\hat{\beta}_3-}\dots \, .
\end{equation}
We can use this relation to obtained very accurately the ratio $\phi_1/\phi_0$, which through eq~\eqref{eq:field_redefinition} gives $\psi'(r=a)/\psi(r=a)$, which we can match with the quantity obtained from the ``in'' solution. 

\subsection{Determining quasinormal mode frequencies}
The methods described in the previous subsections allow two independent determinations of the value $\psi'(a)/\psi(a)$ for the in and up solutions, for every value of the frequency $\omega$. Quasinormal modes correspond to frequencies where both ratios match, and hence the Wronskian~\eqref{eq:wronskian} vanishes. That defines a nonlinear equation for $\omega$. We search numerically for roots of this equation in the complex plane, which correspond to the quasinormal modes of the system. Convergence of the root finder demands an accurate initial guess: we use the values of the $w$-modes for perfect fluid stars as initial guess for low values of the viscosity, and progressively track the evolution of the QNM frequencies for increasing shear viscosity. We have compared the results of our code in the low-viscosity limit against known results for a perfect fluid, finding excellent agreement (see also below). Our code is implemented in \texttt{Mathematica}, and available online~\cite{web:CoG}. 

\section{Results\label{sec:Results}}
\begin{table*}[ht!]
\centering
\small
\setlength{\tabcolsep}{4pt}
\begin{tabular}{lcccc}
\hline\hline
& \multicolumn{4}{c} {$(f \, [\rm kHz], \tau \, [\mu s])$}\\
$\eta_c \,[\mathrm{g\,cm^{-1}\,s^{-1}}]$ & A1 & A2 & B1 & B2 \\
\hline
$3\times10^{29}$ &
(10.4884 , 29.5870) &
(10.4884 , 29.5870) &
(10.4868 , 29.5894) &
(10.4868 , 29.5891) \\
$5\times10^{29}$ &
(10.4795 , 29.6169) &
(10.4794 , 29.6167) &
(10.4769 , 29.6194) &
(10.4768 , 29.6200) \\
$8\times10^{29}$ &
(10.4661 , 29.6619) &
(10.4660 , 29.6608) &
(10.4622 , 29.6639) &
(10.4619 , 29.6658) \\
$1\times10^{30}$ &
(10.4571 , 29.6917) &
(10.4571 , 29.6898) &
(10.4523 , 29.6938) &
(10.4522 , 29.6964) \\
$3\times10^{30}$ &
(10.3692 , 29.9752) &
(10.3705 , 29.9733) &
(10.3564 , 29.9658) &
(10.3539 , 29.9699) \\
$5\times10^{30}$ &
(10.2854 , 30.2463) &
(10.2874 , 30.2467) &
(10.2783 , 30.1107) &
(10.2625 , 30.2600) \\
$8\times10^{30}$ &
(10.1659 , 30.6362) &
(10.1687 , 30.6394) &
(10.1293 , 30.5004) &
(10.1443 , 30.6579) \\
$1\times10^{31}$ &
(10.0898 , 30.8857) &
(10.0932 , 30.8905) &
(10.0608 , 30.7400) &
(10.1271 , 30.8477) \\
\hline\hline
\end{tabular}
\caption{Fundamental quadrupolar spacetime modes written as ($f \, [\rm kHz], \tau \, [\mu s]$), for different values of the central shear viscosity $\eta_c$, and different hydrodynamic frames, as defined in Table~\ref{tab:frames}. In this case, we consider a star with central density $\rho_c = 3 \times 10^{15} \rm g \, cm^{-3}$, and polytropic equation of state with $\kappa=100$, and $n=1$ (leading to a moderate compactness, $M/R=0.21$). We confirm that the $w$-mode frequencies do not depend significantly on the frame choice. The frame B1 shows a larger discrepancy at large viscosities due to mode avoidance, as it comes close to a viscous $\eta$-mode, as we discuss in Section~\ref{ssec:eta_modes}. 
}
\label{tab:modes_frames}
\end{table*}
In this work, we focus on characterising the quadrupolar $\ell=2$ oscillations of viscous stars. First, we test the accuracy and robustness of our framework to determine the axial modes of a viscous star. We have benchmarked our code by comparing the $w$-mode frequencies obtained for very low viscosity against known results in the perfect fluid limit, for a star with $\rho_c=3\times 10^{15}\mathrm{g}/\mathrm{cm}^3$, and polytropic equation of state with $\kappa=100$, and $n=1$, with sub-percent agreement~\cite{kokkotas_private_comm},
and for constant-density stars, finding percent-level agreement for long-lived $w$-modes with Refs.~\cite{Chandrasekhar:1991fi, Kokkotas:1994an}.

 \begin{figure*}[ht!]
    \centering    \includegraphics[width=0.95\linewidth]{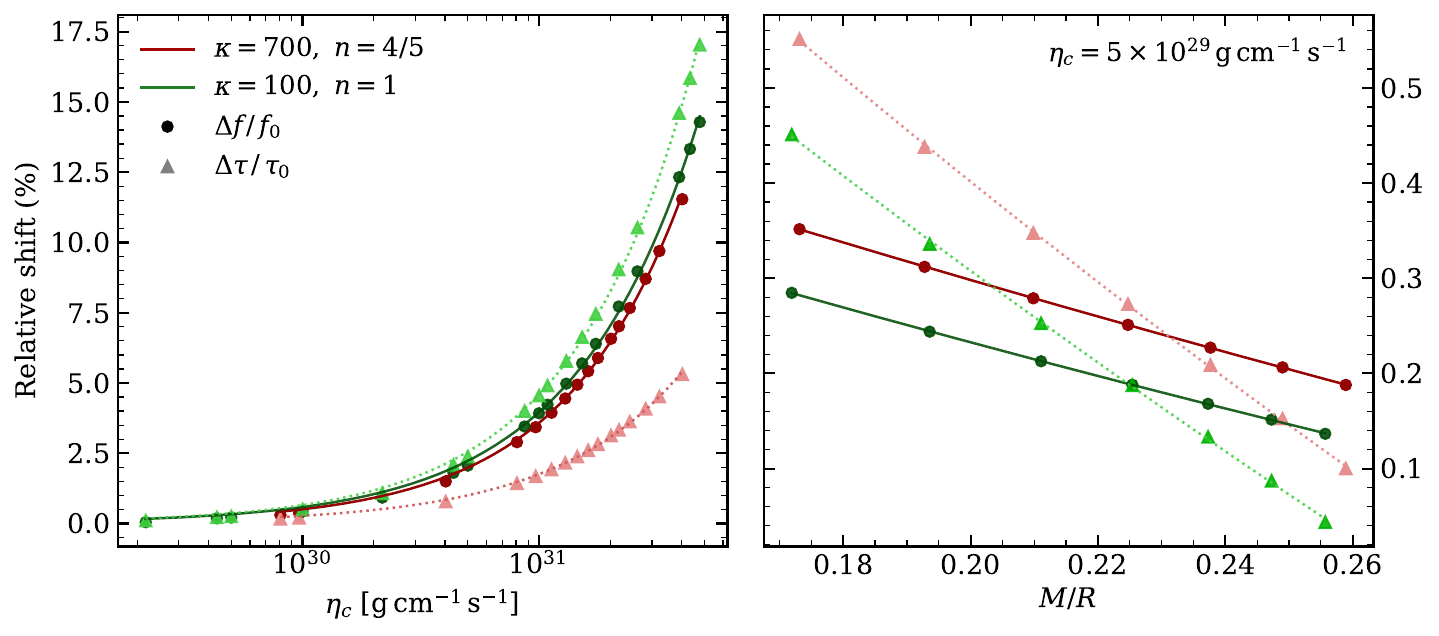}
    \caption{\textbf{Left:} Relative frequency (darker lines, circles) and damping time (lighter lines, triangles) shifts of the quadrupolar $w$-modes as a function of the central shear viscosity of the star, in the A2 hydrodynamic frame. Red (respectively green) lines correspond to two different polytropic equations of state, as indicated by the legend. In both cases we consider a star with central density $\rho_c = 3 \times 10^{15} \mathrm{g/cm^{3}}$. \textbf{Right:} Same as in the left panel, but as a function of the compactness of the star, for a central shear viscosity $\eta_c = 5\times 10^{29}\cgs$.}
    \label{fig:w-modes}
\end{figure*}
Next, we study whether the results we obtain for $w$-modes are independent of the choice of hydrodynamic frame and viscous parametrization. We consider four different parametrizations, summarized in Table~\ref{tab:frames}. We obtain the $w$-mode frequencies for all four frames and different values of the viscosity, and list them in Table~\ref{tab:modes_frames}. we find that all four cases result in an almost identical frequency and damping time for the $w$-mode: the differences between frames are below the percent level. This is strong evidence of frame robustness: as expected, changing the hydrodynamic frame (e.g. changing the value of $\hat{\tau}$) results in a correction to the $w$-mode frequencies which is subleading compared to the difference between the viscous and perfect fluid cases. For this reason, in what follows, we will only report values for one hydrodynamic frame -- the results would be equivalent for all other possible frames. The only difference is whenever mode avoidance occurs, which we discuss below.

\subsection{Characterization of $w$-modes}\label{ssec:w-modes}

We now study how viscosity shifts the oscillation frequency and damping times of $w$-modes. We consider only the A2 parametrization, but as demonstrated in Table~\ref{tab:modes_frames}, the results are largely independent of the hydrodynamic frame. Recall that we write the quasinormal mode frequencies as $\omega = 2\pi f - i/\tau$, with $f,\tau$ denoting the oscillation frequency, and damping time, respectively. Moreover, we use $f_0,\tau_0$ to denote the frequency and damping time of the $w$-mode in the perfect fluid limit, and denote by $\Delta f/f_0 = (f-f_0)/f_0$ the relative frequency shift induced by shear viscosity.

Our results are summarized in Fig.~\ref{fig:w-modes}. The left panel shows the relative shifts in the oscillation frequency and damping time as a function of the central viscosity of the star $\eta_c = \eta(r=0)$. For large values of the shear viscosity, $\eta_c \gtrsim 10^{30}\cgs$, we find frequency and damping time shifts at the percent-level. This magnitude is comparable with the frequency shifts recently reported for radial modes in Ref.~\cite{Keeble:2026bzo}. The dependence on the viscosity is approximately linear, as one could expect from energy estimates, see Ref.~\cite{cutler1987effect}. However, such analysis fails to capture shifts in the oscillation frequencies of the $w$-modes, of a comparable magnitude to those of the damping time.

Figure~\ref{fig:w-modes} (right panel), also shows the effect of compactness on the structure of the $w$ modes. We consider a moderate viscosity configuration, $\eta_c = 5\times 10^{29}\cgs$, and vary the compactness of the star $M/R$. Remarkably we find that the relative frequency and damping time shifts are larger for less compact stars. This can be explained by recalling that $w$-modes are, ultimately, spacetime oscillation modes, and therefore they are most sensitive to the stellar compactness. We find the same scaling with the star's compactness for both equations of state, 
\begin{equation}
    \frac{\Delta f}{f_0} \sim C_f - 1.8\frac{M}{R} \, , \qquad \frac{\Delta \tau}{\tau_0} \sim C_\tau-5.0 \frac{M}{R} \, , 
\end{equation}
where the constants $C_{f,\tau}$ depend on the equation of state, but the slope of the line is the same for both equations of state considered in this work.

\subsection{Characterization of $\eta$-modes}\label{ssec:eta_modes}
\begin{figure}[t!]
    \centering
    \includegraphics[width=0.5\textwidth]{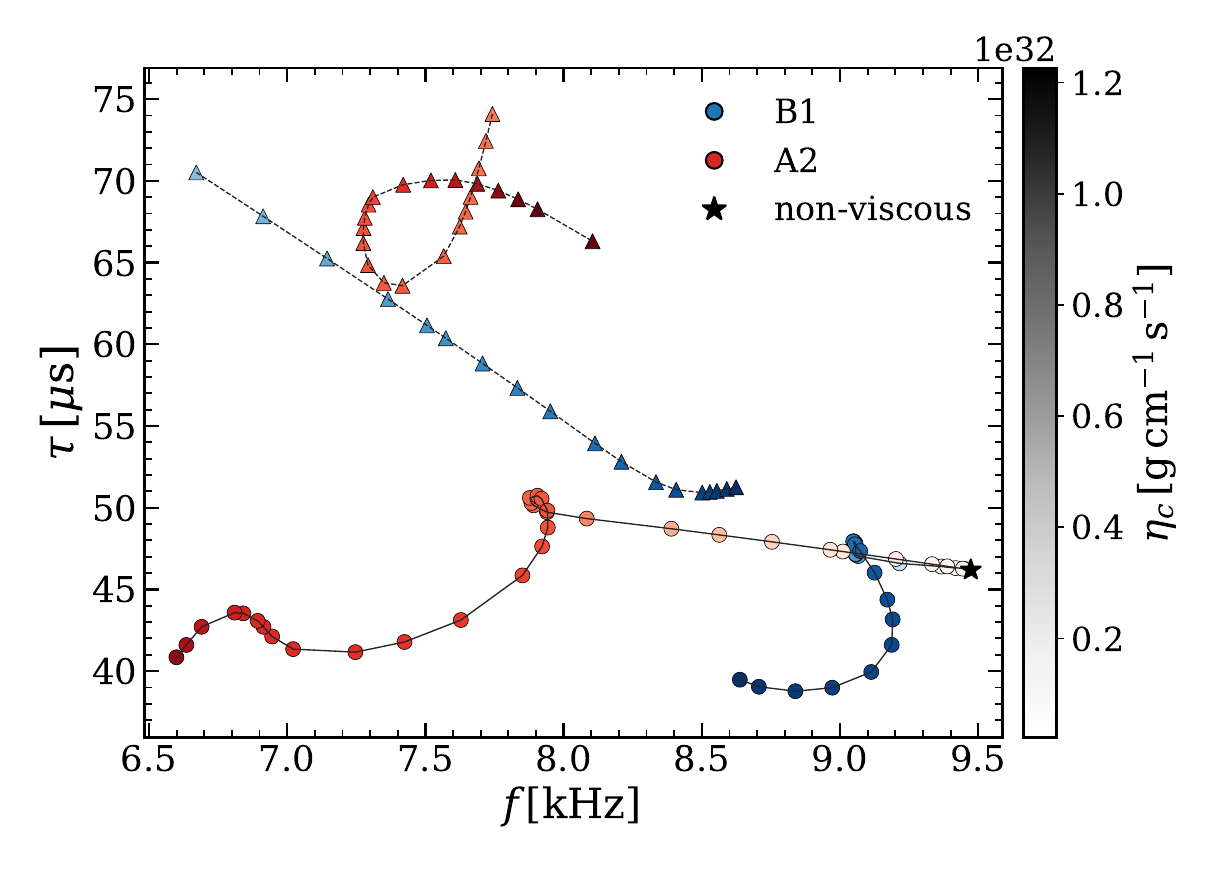}
    \caption{Frequencies (horizontal axis) and damping times (vertical axis) of the quadrupolar oscillation modes as a function of the central viscosity of the star, indicated in the color scale. Lighter colors correspond to less viscous configurations, and the perfect fluid limit is depicted with a star. The blue and red colors correspond to two different hydrodynamic frames. We represent as triangles the $\eta$-modes, which approach and then avoid the family of $w$-modes, illustrating the phenomenon of mode avoidance. 
    }
    \label{fig:eta_modes}
\end{figure}
When studying perturbations of BDNK hydrodynamics, it is natural to expect a new family of modes, associated to shear viscosity. The restoring force of these modes is provided by the transport coefficients, in a combination of the shear viscosity and the regulator $\hat{\tau}$. Therefore, these modes depend more sensitively on the choice of hydrodynamic frame, as opposed to $w$-modes (notice that they do not have an analogue in the perfect fluid limit). Therefore, finding them with a root-finding method becomes a non-trivial task, as we do not have an informative initial guess, beyond the results presented in~\cite{Boyanov:2024jge} in the Cowling approximation, and only for one hydrodynamic frame. Despite this, we have found and characterized the dependence of viscosity of $\eta$-modes for two different parametrizations. Our results are summarized in Fig.~\ref{fig:eta_modes}. 

Fig.~\ref{fig:eta_modes} shows the evolution (in the complex plane) of the quasinormal mode frequencies for increasing values of the central viscosity, as indicated in the color grading. $w$-modes are represented as circles, and they approach the perfect fluid value (indicated with a star) as $\eta_c\to 0$. The shear $\eta$-modes are marked with triangles. We have not been able to track them all the way until very small viscosities, but generically they seem to satisfy $\Im\omega \to 0$ as $\eta_c\to 0$ -- they become undamped, in the perfect fluid limit. Interestingly, these modes also oscillate with $\mathrm{kHz}$ frequencies, and can have damping times in the order of $\mathrm{ms}$, as the $w$-modes.

Mode avoidance is a well-known phenomena, present often when a system has multiple families of modes ($w$- and $\eta $-modes) and free parameters (e.g. compactness of the star, and shear viscosity). This was first discussed decades ago~\cite{vonNeumann}, and is also well-known in the context of asteroseismology~\cite{Christensen:2003}, and in the spectrum of black holes~\cite{Dias:2022oqm, Motohashi:2024fwt, Berti:2025hly}. We also report mode avoidance in the axial oscillations of viscous stars. Fig.~\ref{fig:eta_modes} shows that the families of $w$- and $\eta$-modes approach each other as we increase the shear viscosity, but they never cross, repelling away from each other instead.

We highlight that we provide here the first characterization of this new family of oscillation modes of neutron stars. As their frequencies seem to depend on the details of the hydrodynamic theory under consideration, it would be interesting to investigate their possible excitation either in nonlinear simulations, as well as in other theories such as Israel-Stewart hydrodynamics (a preliminary analysis would seem to indicate that these modes are absent in that case~\cite{Caballero_2025}). Our preliminary results, however, demonstrate (i) that $\eta$-modes can have frequencies and damping times comparable with $w$-modes, and therefore, of relevance in characterizing the dynamics of neutron stars, and (ii) that these modes can produce mode avoidance as they approach the $w$-modes for large values of the shear viscosity, destabilizing the frequencies of the $w$-modes. We highlight that we only find this behavior when increasing the shear viscosity up to values that are possibly beyond what is realistic, even considering the presence of strange matter in the neutron star core. 

\subsection{Viscous damping in ultracompact stars}\label{ssec:ucos}
Finally, we examine the impact of viscosity on the dynamics of ultracompact objects. We consider now constant density stars $\rho=\mathrm{const.}$ -- therefore, we restrict to the hydrodynamic frame B, where there is no direct dependence on the (ill-defined) sound speed. Constant density stars admit solutions with $M/R \leq 4/9$, where the upper limit is the Buchdahl limit~\cite{Buchdahl:1959zz}. In those cases, the surface of the star is \emph{inside} its own lightring, and the potential $V$ has both a maximum (close to $r\sim 3M$), and a minimum (typically inside the star). This minimum signals the appearance of a \emph{stable} lightring, where null radiation (in particular, gravitational waves), can be efficiently trapped. This trapping has been shown to lead to turbulent dynamics~\cite{Benomio:2024lev, Redondo-Yuste:2025hlv, Siemonsen:2025fne, Siemonsen:2025wib}, and even conjectured to lead to nonlinear instabilities~\cite{Keir:2014oka, Benomio:2018ivy, Cardoso:2014sna, Cunha:2017qtt, Cunha:2022gde, Siemonsen:2024snb, Marks:2025jpt, Evstafyeva:2025mvx}. The presence of long-lived modes associated to a stable lightring would introduce echoes in the gravitational wave emission of these objects~\cite{Cardoso:2014sna,Cardoso:2019rvt}, which are currently being searched for in data~\cite{LIGOScientific:2026wpt}. 

\begin{figure}[t]
    \centering
    \includegraphics[width=\columnwidth]{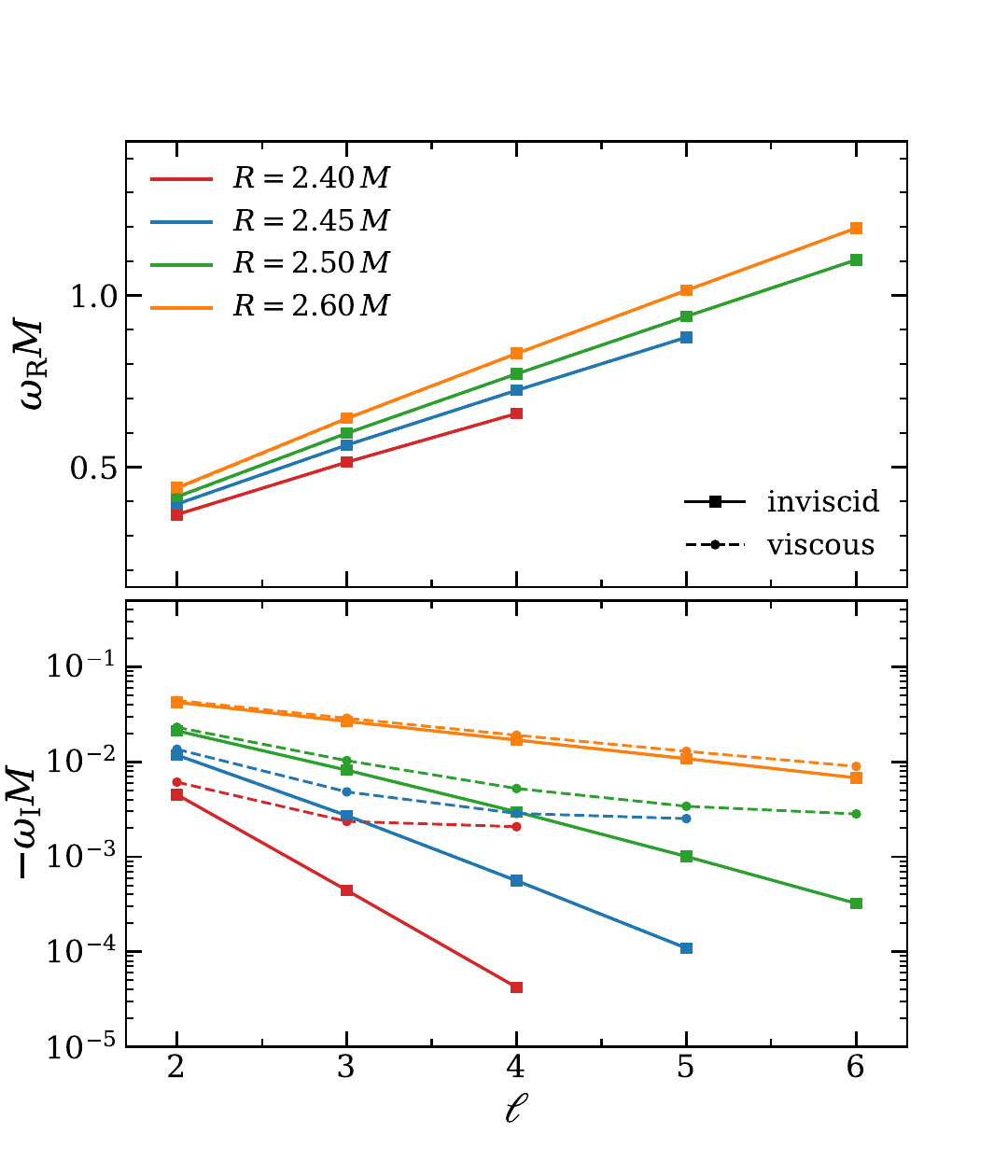}
    \caption{Modes of ultracompact constant-density stars with different compactness. Solid and dotted lines correspond to the ideal and viscous configurations for a central shear viscosity $\eta_c = 10^{31} \rm g \, cm^{-1} \, s^{-1}$, respectively. These have been computed with B parametrization.
    }
    \label{fig:ucos}
\end{figure}
Our results are summarized in Fig.~\ref{fig:ucos}. We consider constant density stars with increasing compactness, in different colors. All these stars are compact enough to have a stable lightring. Fig.~\ref{fig:ucos} shows the real and imaginary parts of the fundamental $w$-mode in that configuration, for different values of the harmonic number $\ell$. In the perfect fluid case, it is known that the frequencies approximately scale as 
\begin{equation}
    \omega_\ell = \Bigl(\ell + \frac{1}{2}\Bigr) \Omega_{\rm LR} - ie^{-\gamma \ell}  \, , 
\end{equation}
where $\Omega_{\rm LR}$ is the frequency of the stable lightring, and $\gamma$ its Lyapunov exponent~\cite{Cardoso:2014sna}. Our results perfectly reproduce this behavior in the inviscid case (solid lines). 

However, when including viscosity (we set here, for simplicity, $\eta_c = 10^{31}\cgs$), the situation is quite different. The oscillation frequencies are unaffected by viscosity -- as we have discussed before, compactness reduces the impact of viscosity on the relative frequency shifts, which are never larger than a few percent for this viscosity. However, the imaginary part, which controls the damping time, changes drastically. In particular, it seems that $|\Im\omega_\ell| \lesssim  10^{-2}$, regardless of the value of $\ell$, and the compactness of the star. Shear viscosity damps the long-lived $w$-modes, with a damping rate that is set by the scale of the transport coefficients, regardless of the compactness of the star. 
Thus, as argued before on qualitative grounds~\cite{Cardoso:2014sna}, and supported by our results, viscosity plays a crucial role in any putative nonlinear instability of compact objects.

Note, however, that when the star becomes genuinely ultracompact $R = 2M(1+\epsilon)$ with $\epsilon \ll 1$, the frequency of the trapped modes also becomes extremely small -- in that case, we do not expect the viscous damping to be efficient~\cite{Boyanov:2024jge}. Therefore, genuinely ultracompact objects may still have long-lived trapped modes, even if they are extremely viscous.

\section{Conclusions}
We started a first-principles investigation of the spectrum of dissipative compact stars. We find that the gravitational $w$-mode oscillation frequencies change by $10\%$ for large shear viscosities $\eta\sim 10^{31}\cgs$. Similarly, viscosity slows down the damping of these modes by a similar amount. For the polytropic equations of state considered in this work, the impact of viscosity is larger for \emph{less} compact stars. We demonstrate that these results do not depend on the additional transport coefficients that the BDNK hydrodynamics theory introduces in order to regulate the causality and stability properties of the theory. 

In other words, gravitational modes are not significantly affected by viscosity. However, we find and characterize a new family of modes with $\sim \mathrm{kHz}$ oscillation frequencies, with no counterpart for a perfect fluid, which we dub $\eta$-modes. These modes depend on the shear viscosity and the \emph{second sound} speed~\cite{Bemfica:2020zjp, Redondo-Yuste:2024vdb, Caballero_2025}, and therefore, they do depend on the additional BDNK transport coefficients. It is important to understand, possibly via time-domain simulations, the actual excitation of these new $\eta$ modes.

Remarkably, for some stellar configurations here considered, we find an \emph{avoided crossing} between $w$- and $\eta$-modes. This phenomenon was first studied by von Neumann and Wigner in the context of atomic physics~\cite{vonNeumann}. Avoided crossings are also present between $p$- and $g$-modes in (Newtonian) stellar oscillations~\cite{Christensen:2003}, and have recently received further attention in the context of black hole physics~\cite{Onozawa:1996ux, Oshita:2021iyn, Motohashi:2024fwt,Oshita:2025ibu, PanossoMacedo:2025xnf}. In this context, an avoided crossing between $\eta$- and $w$-modes renders the spacetime modes very sensitive to the viscous transport coefficients in a small region of parameter space. 

Our study focuses only on the axial sector, the polar sector is expected to yield yet a richer structure. In particular, the impact of dissipation on other (realistic) families of modes, such as f-, p- and g-modes is yet to be studied. Dissipative effects will also affect the r-modes of rotating stars, possibly quenching instabilities~\cite{Kojima:1992ie, Andersson:1997xt, Andersson:2000mf, Ruoff:2001fq, Stavridis:2004hg}. We leave an exploration of the impact of shear viscosity on r-modes extending those of Refs.~\cite{Pons:2005gb,Redondo-Yuste:2025ktt} for future work.

\section*{Acknowledgements}
We are indebted to Kostas Kokkotas for sharing results on the $w$-modes for inviscid stars and comments, and to Caio Macedo for valuable discussions on determining QNM frequencies, and comments on a previous version of this manuscript. 
The Center of Gravity is a Center of Excellence funded by the Danish National Research Foundation under grant No. DNRF184.
We acknowledge support by VILLUM Foundation (grant no. VIL37766).
V.C.\ is a Villum Investigator.  
V.C. acknowledges financial support provided under the European Union’s H2020 ERC Advanced Grant “Black holes: gravitational engines of discovery” grant agreement no. Gravitas–101052587. 
Views and opinions expressed are however those of the author only and do not necessarily reflect those of the European Union or the European Research Council. Neither the European Union nor the granting authority can be held responsible for them.
This project has received funding from the European Union's Horizon 2020 research and innovation programme under the Marie Sklodowska-Curie grant agreement No 101007855 and No 101131233.
This work is supported by Simons Foundation International \cite{sfi} and the Simons Foundation \cite{sf} through Simons Foundation grant SFI-MPS-BH-00012593-11.
J.~R.-Y. is supported by NSF Grants No.~AST-2307146, No.~PHY-2513337, No.~PHY-090003, and No.~PHY-20043, by NASA Grant No.~21-ATP21-0010, by John Templeton Foundation Grant No.~62840, by the Simons Foundation [MPS-SIP-00001698, E.B.], by the Simons Foundation International [SFI-MPS-BH-00012593-02], and by Italian Ministry of Foreign Affairs and International Cooperation Grant No.~PGR01167.
J.J.O.G acknowledges support by the Spanish Ministry of Science, Innovation and Universities under the FPU predoctoral grant FPU24/02041
S.B. is supported by the Universitat de les Illes Balears (UIB) with funds from the Programa de Foment de la Recerca i la Innovació de la UIB 2024-2026 (supported by the yearly plan of the Tourist Stay Tax ITS2023-086); the Spanish Agencia Estatal de Investigación grants PID2022-138626NB-I00, RED2024-153978-E, RED2024-153735-E, funded by MICIU/AEI/10.13039/501100011033 and the ERDF/EU; and the Comunitat Autònoma de les Illes Balears through the Conselleria d'Educació i Universitats with funds from the ERDF (SINCO2022/18146 - Plataforma HiTech-IAC3-BIO).

\bibliography{biblio}

@misc{sfi,
    howpublished = "\url{https://www.sfi.org.bm/}",
    author = "SFI"
}

@misc{sf,
    howpublished = "\url{https://www.simonsfoundation.org/}",
    author = "SF"
}

@misc{kokkotas_private_comm,
  author       = {Kokkotas, Kostas},
  title        = {},
  howpublished = {},
  year         = {2026},
  note         = {Private communication}
}

@article{Kokkotas:1994an,
    author = "Kokkotas, K. D.",
    title = "{Axial modes for relativistic stars}",
    journal = "Mon. Not. Roy. Astron. Soc.",
    volume = "268",
    pages = "1015",
    year = "1994"
}

@article{HegadeKR:2026iou,
    author = "Hegade K. R., Abhishek and Yang, Yumu and Hippert, Mauricio and Noronha-Hostler, Jacquelyn and Noronha, Jorge and Yunes, Nicol{\'a}s",
    title = "{Dynamical tidal response of neutron stars as a probe of dense-matter properties}",
    eprint = "2603.26886",
    archivePrefix = "arXiv",
    primaryClass = "gr-qc",
    month = "3",
    year = "2026"
}

@article{Stavridis:2004hg,
    author = "Stavridis, Adamantios and Kokkotas, Kostas D.",
    title = "{Evolution equations for slowly rotating stars}",
    eprint = "gr-qc/0411019",
    archivePrefix = "arXiv",
    doi = "10.1142/S021827180500592X",
    journal = "Int. J. Mod. Phys. D",
    volume = "14",
    pages = "543",
    year = "2005"
}

@article{Chandrasekhar:1991fi,
    author = "Chandrasekhar, Subrahmanyan and Ferrari, V.",
    editor = "Wali, K. C.",
    title = "{On the non-radial oscillations of a star}",
    doi = "10.1098/rspa.1991.0016",
    journal = "Proc. Roy. Soc. Lond. A",
    volume = "432",
    pages = "247--279",
    year = "1991"
}

@BOOK{Tolman_RTC,
       author = {{Tolman}, R.~C.},
        title = "{Relativity, Thermodynamics, and Cosmology}",
         year = 1934,
       adsurl = {https://ui.adsabs.harvard.edu/abs/1934rtc..book.....T}
}

@article{Caballero_2025,
    author = "Caballero, Daniel A. and Yunes, Nicol{\'a}s",
    title = "{Neutron star radial perturbations for causal, viscous, relativistic fluids}",
    eprint = "2506.09149",
    archivePrefix = "arXiv",
    primaryClass = "gr-qc",
    doi = "10.1103/cl4s-n7nr",
    journal = "Phys. Rev. D",
    volume = "112",
    number = "6",
    pages = "063050",
    year = "2025"
}

@article{Bemfica_2018,
   title={Causality and existence of solutions of relativistic viscous fluid dynamics with gravity},
   volume={98},
   ISSN={2470-0029},
   url={http://dx.doi.org/10.1103/PhysRevD.98.104064},
   DOI={10.1103/physrevd.98.104064},
   number={10},
   journal={Physical Review D},
   publisher={American Physical Society (APS)},
   author={Bemfica, Fábio S. and Disconzi, Marcelo M. and Noronha, Jorge},
   year={2018},
   month={Nov}
}

@article{Bemfica_2019,
  title = {Nonlinear causality of general first-order relativistic viscous hydrodynamics},
  author = {Bemfica, F\'abio S. and Disconzi, Marcelo M. and Noronha, Jorge},
  journal = {Phys. Rev. D},
  volume = {100},
  issue = {10},
  pages = {104020},
  numpages = {13},
  year = {2019},
  month = {Nov},
  publisher = {American Physical Society},
  doi = {10.1103/PhysRevD.100.104020},
  url = {https://link.aps.org/doi/10.1103/PhysRevD.100.104020}
}

@article{Kovtun_2019,
   title={First-order relativistic hydrodynamics is stable},
   volume={2019},
   ISSN={1029-8479},
   url={http://dx.doi.org/10.1007/JHEP10(2019)034},
   DOI={10.1007/jhep10(2019)034},
   number={10},
   journal={Journal of High Energy Physics},
   publisher={Springer Science and Business Media LLC},
   author={Kovtun, Pavel},
   year={2019},
   month={Oct}
}

@misc{web:CoG,
  note = "{Check the Center of Gravity webpage for publicly available material:
  \\ \url{https://the-center-of-gravity.com/}
  }"
  }

@article{Kojima:1992ie,
    author = "Kojima, Y.",
    title = "{Equations governing the nonradial oscillations of a slowly rotating relativistic star}",
    doi = "10.1103/PhysRevD.46.4289",
    journal = "Phys. Rev. D",
    volume = "46",
    pages = "4289--4303",
    year = "1992"
}

@article{Pons:2005gb,
    author = "Pons, J. A. and Gualtieri, L. and Miralles, J. A. and Ferrari, V.",
    title = "{Relativistic r-modes and shear viscosity: Regularizing the continuous spectrum}",
    eprint = "astro-ph/0504062",
    archivePrefix = "arXiv",
    doi = "10.1111/j.1365-2966.2005.09429.x",
    journal = "Mon. Not. Roy. Astron. Soc.",
    volume = "363",
    pages = "121--130",
    year = "2005"
}

@inproceedings{Chabanov:2024yqv,
    author = "Chabanov, Michail and Cruz-Osorio, Alejandro and Ecker, Christian and Meringolo, Claudio and Musolino, Carlo and Rezzolla, Luciano and Tootle, Samuel and Topolski, Konrad",
    title = "{Microphysical Aspects of Binary Neutron Star Mergers}",
    booktitle = "{High Performance Computing in Science and Engineering '22}",
    doi = "10.1007/978-3-031-46870-4\_2",
    year = "2024"
}

@article{Chabanov:2023abq,
    author = "Chabanov, Michail and Rezzolla, Luciano",
    title = "{Numerical modeling of bulk viscosity in neutron stars}",
    eprint = "2311.13027",
    archivePrefix = "arXiv",
    primaryClass = "gr-qc",
    doi = "10.1103/PhysRevD.111.044074",
    journal = "Phys. Rev. D",
    volume = "111",
    number = "4",
    pages = "044074",
    year = "2025"
}

@article{Chabanov:2023blf,
    author = "Chabanov, Michail and Rezzolla, Luciano",
    title = "{Impact of Bulk Viscosity on the Postmerger Gravitational-Wave Signal from Merging Neutron Stars}",
    eprint = "2307.10464",
    archivePrefix = "arXiv",
    primaryClass = "gr-qc",
    doi = "10.1103/PhysRevLett.134.071402",
    journal = "Phys. Rev. Lett.",
    volume = "134",
    number = "7",
    pages = "071402",
    year = "2025"
}

@article{Pandya:2022sff,
    author = "Pandya, Alex and Most, Elias R. and Pretorius, Frans",
    title = "{Causal, stable first-order viscous relativistic hydrodynamics with ideal gas microphysics}",
    eprint = "2209.09265",
    archivePrefix = "arXiv",
    primaryClass = "gr-qc",
    doi = "10.1103/PhysRevD.106.123036",
    journal = "Phys. Rev. D",
    volume = "106",
    number = "12",
    pages = "123036",
    year = "2022"
}

@article{Most:2022yhe,
    author = "Most, Elias R. and Haber, Alexander and Harris, Steven P. and Zhang, Ziyuan and Alford, Mark G. and Noronha, Jorge",
    title = "{Emergence of Microphysical Bulk Viscosity in Binary Neutron Star Postmerger Dynamics}",
    eprint = "2207.00442",
    archivePrefix = "arXiv",
    primaryClass = "astro-ph.HE",
    reportNumber = "INT-PUB-22-017",
    doi = "10.3847/2041-8213/ad454f",
    journal = "Astrophys. J. Lett.",
    volume = "967",
    number = "1",
    pages = "L14",
    year = "2024"
}

@article{Pandya:2022pif,
    author = "Pandya, Alex and Most, Elias R. and Pretorius, Frans",
    title = "{Conservative finite volume scheme for first-order viscous relativistic hydrodynamics}",
    eprint = "2201.12317",
    archivePrefix = "arXiv",
    primaryClass = "gr-qc",
    doi = "10.1103/PhysRevD.105.123001",
    journal = "Phys. Rev. D",
    volume = "105",
    number = "12",
    pages = "123001",
    year = "2022"
}

@article{Most:2021zvc,
    author = "Most, Elias R. and Harris, Steven P. and Plumberg, Christopher and Alford, Mark G. and Noronha, Jorge and Noronha-Hostler, Jacquelyn and Pretorius, Frans and Witek, Helvi and Yunes, Nicol\'as",
    title = "{Projecting the likely importance of weak-interaction-driven bulk viscosity in neutron star mergers}",
    eprint = "2107.05094",
    archivePrefix = "arXiv",
    primaryClass = "astro-ph.HE",
    reportNumber = "INT-PUB-21-015",
    doi = "10.1093/mnras/stab2793",
    journal = "Mon. Not. Roy. Astron. Soc.",
    volume = "509",
    number = "1",
    pages = "1096--1108",
    year = "2021"
}

@article{HegadeKR:2024slr,
    author = "Hegade K. R., Abhishek and Ripley, Justin L. and Yunes, Nicol\'as",
    title = "{Dissipative tidal effects to next-to-leading order and constraints on the dissipative tidal deformability using gravitational wave data}",
    eprint = "2407.02584",
    archivePrefix = "arXiv",
    primaryClass = "gr-qc",
    doi = "10.1103/PhysRevD.110.044041",
    journal = "Phys. Rev. D",
    volume = "110",
    number = "4",
    pages = "044041",
    year = "2024"
}

@article{HegadeKR:2024agt,
    author = "Hegade K. R., Abhishek and Ripley, Justin L. and Yunes, Nicol\'as",
    title = "{Dynamical tidal response of nonrotating relativistic stars}",
    eprint = "2403.03254",
    archivePrefix = "arXiv",
    primaryClass = "gr-qc",
    doi = "10.1103/PhysRevD.109.104064",
    journal = "Phys. Rev. D",
    volume = "109",
    number = "10",
    pages = "104064",
    year = "2024"
}

@article{Ripley:2023qxo,
    author = "Ripley, Justin L. and Hegade K. R., Abhishek and Yunes, Nicolas",
    title = "{Probing internal dissipative processes of neutron stars with gravitational waves during the inspiral of neutron star binaries}",
    eprint = "2306.15633",
    archivePrefix = "arXiv",
    primaryClass = "gr-qc",
    doi = "10.1103/PhysRevD.108.103037",
    journal = "Phys. Rev. D",
    volume = "108",
    number = "10",
    pages = "103037",
    year = "2023"
}

@article{Ripley:2023lsq,
    author = "Ripley, Justin L. and Hegade K. R., Abhishek and Chandramouli, Rohit S. and Yunes, Nicolas",
    title = "{A constraint on the dissipative tidal deformability of neutron stars}",
    eprint = "2312.11659",
    archivePrefix = "arXiv",
    primaryClass = "gr-qc",
    doi = "10.1038/s41550-024-02323-7",
    journal = "Nature Astron.",
    volume = "8",
    number = "10",
    pages = "1277--1283",
    year = "2024"
}

@article{Alford:2018lhf,
    author = "Alford, Mark G. and Harris, Steven P.",
    title = "{Beta equilibrium in neutron star mergers}",
    eprint = "1803.00662",
    archivePrefix = "arXiv",
    primaryClass = "nucl-th",
    doi = "10.1103/PhysRevC.98.065806",
    journal = "Phys. Rev. C",
    volume = "98",
    number = "6",
    pages = "065806",
    year = "2018"
}

@article{Alford:2020lla,
    author = "Alford, Mark and Harutyunyan, Arus and Sedrakian, Armen",
    title = "{Bulk Viscous Damping of Density Oscillations in Neutron Star Mergers}",
    eprint = "2006.07975",
    archivePrefix = "arXiv",
    primaryClass = "nucl-th",
    doi = "10.3390/particles3020034",
    journal = "Particles",
    volume = "3",
    number = "2",
    pages = "500--517",
    year = "2020"
}

@article{Benomio:2024lev,
    author = "Benomio, Gabriele and C\'ardenas-Avenda\~no, Alejandro and Pretorius, Frans and Sullivan, Andrew",
    title = "{On turbulence for spacetimes with stable trapping}",
    eprint = "2411.17445",
    archivePrefix = "arXiv",
    primaryClass = "gr-qc",
    month = "11",
    year = "2024"
}

@article{Kovtun:2004de,
    author = "Kovtun, P. and Son, Dan T. and Starinets, Andrei O.",
    title = "{Viscosity in strongly interacting quantum field theories from black hole physics}",
    eprint = "hep-th/0405231",
    archivePrefix = "arXiv",
    reportNumber = "INT-PUB-04-09, UW-PT-04-04",
    doi = "10.1103/PhysRevLett.94.111601",
    journal = "Phys. Rev. Lett.",
    volume = "94",
    pages = "111601",
    year = "2005"
}

@article{Cunha:2017qtt,
    author = "Cunha, Pedro V. P. and Berti, Emanuele and Herdeiro, Carlos A. R.",
    title = "{Light-Ring Stability for Ultracompact Objects}",
    eprint = "1708.04211",
    archivePrefix = "arXiv",
    primaryClass = "gr-qc",
    doi = "10.1103/PhysRevLett.119.251102",
    journal = "Phys. Rev. Lett.",
    volume = "119",
    number = "25",
    pages = "251102",
    year = "2017"
}

@article{Kovtun:2003wp,
    author = "Kovtun, Pavel and Son, Dam T. and Starinets, Andrei O.",
    title = "{Holography and hydrodynamics: Diffusion on stretched horizons}",
    eprint = "hep-th/0309213",
    archivePrefix = "arXiv",
    reportNumber = "UW-PT-03-21, INT-PUB-03-17",
    doi = "10.1088/1126-6708/2003/10/064",
    journal = "JHEP",
    volume = "10",
    pages = "064",
    year = "2003"
}

@article{Boyanov:2024jge,
    author = "Boyanov, Valentin and Cardoso, Vitor and Kokkotas, Kostas D. and Redondo-Yuste, Jaime",
    title = "{The dynamical response of viscous objects to gravitational waves}",
    eprint = "2411.16861",
    archivePrefix = "arXiv",
    primaryClass = "gr-qc",
    month = "11",
    year = "2024",
    journal = ""
}

@article{Redondo-Yuste:2024vdb,
    author = "Redondo-Yuste, Jaime",
    title = "{Perturbations of relativistic dissipative stars}",
    eprint = "2411.16841",
    archivePrefix = "arXiv",
    primaryClass = "gr-qc",
    month = "11",
    year = "2024",
    journal= ""
}

@MISC{1967ApJ...149..591T,
       author = {{Thorne}, Kip S. and {Campolattaro}, Alfonso},
        title = "{Non-Radial Pulsation of General-Relativistic Stellar Models. I. Analytic Analysis for L >= 2}",
         year = 1967,
        month = sep,
        pages = {591},
          doi = {10.1086/149288},
    publisher = {IOP},
       adsurl = {https://ui.adsabs.harvard.edu/abs/1967ApJ...149..591T}
}

@article{Ruoff:2001fq,
    author = "Ruoff, Johannes and Stavridis, Adamantios and Kokkotas, Kostas D.",
    title = "{Evolution equations for the perturbations of slowly rotating relativistic stars}",
    eprint = "gr-qc/0109065",
    archivePrefix = "arXiv",
    doi = "10.1046/j.1365-8711.2002.05329.x",
    journal = "Mon. Not. Roy. Astron. Soc.",
    volume = "332",
    pages = "676",
    year = "2002"
}

@article{Andersson:1997xt,
    author = "Andersson, Nils",
    title = "{A New class of unstable modes of rotating relativistic stars}",
    eprint = "gr-qc/9706075",
    archivePrefix = "arXiv",
    doi = "10.1086/305919",
    journal = "Astrophys. J.",
    volume = "502",
    pages = "708--713",
    year = "1998"
}

@article{Andersson:2000mf,
    author = "Andersson, Nils and Kokkotas, Kostas D.",
    title = "{The R mode instability in rotating neutron stars}",
    eprint = "gr-qc/0010102",
    archivePrefix = "arXiv",
    doi = "10.1142/S0218271801001062",
    journal = "Int. J. Mod. Phys. D",
    volume = "10",
    pages = "381--442",
    year = "2001"
}

@article{Kokkotas:1999bd,
    author = "Kokkotas, Kostas D. and Schmidt, Bernd G.",
    title = "{Quasinormal modes of stars and black holes}",
    eprint = "gr-qc/9909058",
    archivePrefix = "arXiv",
    doi = "10.12942/lrr-1999-2",
    journal = "Living Rev. Rel.",
    volume = "2",
    pages = "2",
    year = "1999"
}

@article{Bemfica:2020zjp,
    author = "Bemfica, Fabio S. and Disconzi, Marcelo M. and Noronha, Jorge",
    title = "{First-Order General-Relativistic Viscous Fluid Dynamics}",
    eprint = "2009.11388",
    archivePrefix = "arXiv",
    primaryClass = "gr-qc",
    doi = "10.1103/PhysRevX.12.021044",
    journal = "Phys. Rev. X",
    volume = "12",
    number = "2",
    pages = "021044",
    year = "2022"
}

@article{Bemfica:2019knx,
    author = "Bemfica, F\'abio S. and Bemfica, F\'abio S. and Disconzi, Marcelo M. and Disconzi, Marcelo M. and Noronha, Jorge and Noronha, Jorge",
    title = "{Nonlinear Causality of General First-Order Relativistic Viscous Hydrodynamics}",
    eprint = "1907.12695",
    archivePrefix = "arXiv",
    primaryClass = "gr-qc",
    doi = "10.1103/PhysRevD.100.104020",
    journal = "Phys. Rev. D",
    volume = "100",
    number = "10",
    pages = "104020",
    year = "2019",
    note = "[Erratum: Phys.Rev.D 105, 069902 (2022)]"
}

@article{Bemfica:2017wps,
    author = "Bemfica, F\'abio S. and Disconzi, Marcelo M. and Noronha, Jorge",
    title = "{Causality and existence of solutions of relativistic viscous fluid dynamics with gravity}",
    eprint = "1708.06255",
    archivePrefix = "arXiv",
    primaryClass = "gr-qc",
    doi = "10.1103/PhysRevD.98.104064",
    journal = "Phys. Rev. D",
    volume = "98",
    number = "10",
    pages = "104064",
    year = "2018"
}

@article{Kovtun:2012rj,
    author = "Kovtun, Pavel",
    title = "{Lectures on hydrodynamic fluctuations in relativistic theories}",
    eprint = "1205.5040",
    archivePrefix = "arXiv",
    primaryClass = "hep-th",
    doi = "10.1088/1751-8113/45/47/473001",
    journal = "J. Phys. A",
    volume = "45",
    pages = "473001",
    year = "2012"
}

@article{Keir:2014oka,
    author = "Keir, Joe",
    title = "{Slowly decaying waves on spherically symmetric spacetimes and ultracompact neutron stars}",
    eprint = "1404.7036",
    archivePrefix = "arXiv",
    primaryClass = "gr-qc",
    doi = "10.1088/0264-9381/33/13/135009",
    journal = "Class. Quant. Grav.",
    volume = "33",
    number = "13",
    pages = "135009",
    year = "2016"
}

@article{Cardoso:2014sna,
    author = "Cardoso, Vitor and Crispino, Lu\'\i{}s C. B. and Macedo, Caio F. B. and Okawa, Hirotada and Pani, Paolo",
    title = "{Light rings as observational evidence for event horizons: long-lived modes, ergoregions and nonlinear instabilities of ultracompact objects}",
    eprint = "1406.5510",
    archivePrefix = "arXiv",
    primaryClass = "gr-qc",
    doi = "10.1103/PhysRevD.90.044069",
    journal = "Phys. Rev. D",
    volume = "90",
    number = "4",
    pages = "044069",
    year = "2014"
}

@article{Redondo-Yuste:2025hlv,
    author = "Redondo-Yuste, Jaime and C\'ardenas-Avenda\~no, Alejandro",
    title = "{Perturbative and non-linear analyses of gravitational turbulence in spacetimes with stable light rings}",
    eprint = "2502.18643",
    archivePrefix = "arXiv",
    primaryClass = "gr-qc",
    month = "2",
    year = "2025"
}

@article{Marks:2025jpt,
    author = "Marks, Gareth Arturo and Staelens, Seppe J. and Evstafyeva, Tamara and Sperhake, Ulrich",
    title = "{Long-term stable nonlinear evolutions of ultracompact black-hole mimickers}",
    eprint = "2504.17775",
    archivePrefix = "arXiv",
    primaryClass = "gr-qc",
    month = "4",
    year = "2025"
}

@book{Landau:1987,
  title={Fluid Mechanics: Volume 6},
  author={Landau, Lev Davidovich and Lifshitz, Evgenii Mikhailovich},
  volume={6},
  year={1987},
  publisher={Elsevier}
}

@article{Hiscock:1983,
    author = "Hiscock, W. A. and Lindblom, L.",
    title = "{Stability and causality in dissipative relativistic fluids}",
    doi = "10.1016/0003-4916(83)90288-9",
    journal = "Annals Phys.",
    volume = "151",
    pages = "466--496",
    year = "1983"
}

@article{Hiscock:1985,
    author = "Hiscock, William A. and Lindblom, Lee",
    title = "{Generic instabilities in first-order dissipative relativistic fluid theories}",
    doi = "10.1103/PhysRevD.31.725",
    journal = "Phys. Rev. D",
    volume = "31",
    pages = "725--733",
    year = "1985"
}

@article{Ghosh:2025wfx,
    author = "Ghosh, Suprovo and Hern{\'a}ndez, Jos{\'e} Luis and Pradhan, Bikram Keshari and Manuel, Cristina and Chatterjee, Debarati and Tolos, Laura",
    title = "{Tidal heating in binary inspiral of strange quark stars}",
    eprint = "2504.07659",
    archivePrefix = "arXiv",
    primaryClass = "gr-qc",
    doi = "10.1103/h9r7-ld41",
    journal = "Phys. Rev. D",
    volume = "112",
    number = "8",
    pages = "084072",
    year = "2025"
}

@article{Ghosh:2025glz,
    author = "Ghosh, Suprovo and Mukherjee, Samanwaya and Bose, Sukanta and Chatterjee, Debarati",
    title = "{Tidal dissipation in binary neutron star inspirals from hyperon bulk viscosity: Phase modeling and parameter estimation bias}",
    eprint = "2503.14606",
    archivePrefix = "arXiv",
    primaryClass = "gr-qc",
    reportNumber = "LIGO-P2500093",
    doi = "10.1093/mnras/staf1652",
    journal = "Mon. Not. Roy. Astron. Soc.",
    volume = "2987",
    pages = "2996",
    year = "2025"
}

@article{Shum:2025jnl,
    author = "Shum, Harry L. H. and Abalos, Fernando and Bea, Yago and Bezares, Miguel and Figueras, Pau and Palenzuela, Carlos",
    title = "{Neutron star evolution with BDNK viscous hydrodynamics framework}",
    eprint = "2509.15303",
    archivePrefix = "arXiv",
    primaryClass = "gr-qc",
    month = "9",
    year = "2025"
}

@article{Bea:2025eov,
    author = "Bea, Yago",
    title = "{Relativistic Navier-Stokes description of the quark-gluon plasma radial flow}",
    eprint = "2503.02931",
    archivePrefix = "arXiv",
    primaryClass = "hep-ph",
    month = "3",
    year = "2025"
}

@article{Mendes:2025oib,
    author = "Mendes, Raissa F. P. and Guerrieri, Amanda and Muniz, Jo{\~a}o V. M. and Rocha, Gabriel S. and Denicol, Gabriel S.",
    title = "{Radial oscillations of viscous neutron stars: Zero diffusion case}",
    eprint = "2509.12330",
    archivePrefix = "arXiv",
    primaryClass = "gr-qc",
    doi = "10.1103/fs15-pj7m",
    journal = "Phys. Rev. D",
    volume = "113",
    number = "2",
    pages = "024010",
    year = "2026"
}

@article{Baym:2017whm,
    author = "Baym, Gordon and Hatsuda, Tetsuo and Kojo, Toru and Powell, Philip D. and Song, Yifan and Takatsuka, Tatsuyuki",
    title = "{From hadrons to quarks in neutron stars: a review}",
    eprint = "1707.04966",
    archivePrefix = "arXiv",
    primaryClass = "astro-ph.HE",
    reportNumber = "RIKEN-ITHEMS-REPORT-17, RIKEN-QHP-316, RIKEN-iTHEMS-Report-17",
    doi = "10.1088/1361-6633/aaae14",
    journal = "Rept. Prog. Phys.",
    volume = "81",
    number = "5",
    pages = "056902",
    year = "2018"
}

@article{LIGOScientific:2017vwq,
    author = "Abbott, B. P. and others",
    collaboration = "LIGO Scientific, Virgo",
    title = "{GW170817: Observation of Gravitational Waves from a Binary Neutron Star Inspiral}",
    eprint = "1710.05832",
    archivePrefix = "arXiv",
    primaryClass = "gr-qc",
    reportNumber = "LIGO-P170817",
    doi = "10.1103/PhysRevLett.119.161101",
    journal = "Phys. Rev. Lett.",
    volume = "119",
    number = "16",
    pages = "161101",
    year = "2017"
}

@article{LIGOScientific:2018hze,
    author = "Abbott, B. P. and others",
    collaboration = "LIGO Scientific, Virgo",
    title = "{Properties of the binary neutron star merger GW170817}",
    eprint = "1805.11579",
    archivePrefix = "arXiv",
    primaryClass = "gr-qc",
    doi = "10.1103/PhysRevX.9.011001",
    journal = "Phys. Rev. X",
    volume = "9",
    number = "1",
    pages = "011001",
    year = "2019"
}

@article{ET:2019dnz,
    author = "Maggiore, Michele and others",
    collaboration = "ET",
    title = "{Science Case for the Einstein Telescope}",
    eprint = "1912.02622",
    archivePrefix = "arXiv",
    primaryClass = "astro-ph.CO",
    doi = "10.1088/1475-7516/2020/03/050",
    journal = "JCAP",
    volume = "03",
    pages = "050",
    year = "2020"
}

@article{ET:2025xjr,
    author = "Abac, Adrian and others",
    collaboration = "ET",
    title = "{The Science of the Einstein Telescope}",
    eprint = "2503.12263",
    archivePrefix = "arXiv",
    primaryClass = "gr-qc",
    reportNumber = "ET-0036C-25",
    month = "3",
    year = "2025"
}

@article{Reitze:2019iox,
    author = "Reitze, David and others",
    title = "{Cosmic Explorer: The U.S. Contribution to Gravitational-Wave Astronomy beyond LIGO}",
    eprint = "1907.04833",
    archivePrefix = "arXiv",
    primaryClass = "astro-ph.IM",
    reportNumber = "LIGO-P1900316",
    journal = "Bull. Am. Astron. Soc.",
    volume = "51",
    number = "7",
    pages = "035",
    year = "2019"
}

@article{cutler1987effect,
  title={The effect of viscosity on neutron star oscillations},
  author={Cutler, Curt and Lindblom, Lee},
  journal={Astrophysical Journal, Part 1 (ISSN 0004-637X), vol. 314, March 1, 1987, p. 234-241.},
  volume={314},
  pages={234--241},
  year={1987}
}

@article{cutler1990damping,
  title={Damping times for neutron star oscillations},
  author={Cutler, Curt and Lindblom, Lee and Splinter, Randall J},
  journal={Astrophysical Journal, Part 1 (ISSN 0004-637X), vol. 363, Nov. 10, 1990, p. 603-611.},
  volume={363},
  pages={603--611},
  year={1990}
}

@article{Alford:2017rxf,
    author = "Alford, Mark G. and Bovard, Luke and Hanauske, Matthias and Rezzolla, Luciano and Schwenzer, Kai",
    title = "{Viscous Dissipation and Heat Conduction in Binary Neutron-Star Mergers}",
    eprint = "1707.09475",
    archivePrefix = "arXiv",
    primaryClass = "gr-qc",
    doi = "10.1103/PhysRevLett.120.041101",
    journal = "Phys. Rev. Lett.",
    volume = "120",
    number = "4",
    pages = "041101",
    year = "2018"
}

@article{Ghosh:2023vrx,
    author = "Ghosh, Suprovo and Pradhan, Bikram Keshari and Chatterjee, Debarati",
    title = "{Tidal heating as a direct probe of strangeness inside neutron stars}",
    eprint = "2306.14737",
    archivePrefix = "arXiv",
    primaryClass = "gr-qc",
    reportNumber = "LIGO-P2300188",
    doi = "10.1103/PhysRevD.109.103036",
    journal = "Phys. Rev. D",
    volume = "109",
    number = "10",
    pages = "103036",
    year = "2024"
}

@article{Redondo-Yuste:2025ktt,
    author = "Redondo-Yuste, Jaime and Cardoso, Vitor",
    title = "{Superradiant amplification by rotating viscous compact objects}",
    eprint = "2506.13850",
    archivePrefix = "arXiv",
    primaryClass = "gr-qc",
    doi = "10.1103/gvd3-lqkv",
    journal = "Phys. Rev. D",
    volume = "112",
    number = "6",
    pages = "L061501",
    year = "2025"
}

@article{Keeble:2026bzo,
    author = "Keeble, Lennox S. and Redondo-Yuste, Jaime",
    title = "{Radial Oscillations of Viscous Stars}",
    eprint = "2603.23622",
    archivePrefix = "arXiv",
    primaryClass = "gr-qc",
    month = "3",
    year = "2026"
}

@article{LIGOScientific:2017ync,
    author = "Abbott, B. P. and others",
    collaboration = "LIGO Scientific, Virgo, Fermi GBM, INTEGRAL, IceCube, AstroSat Cadmium Zinc Telluride Imager Team, IPN, Insight-Hxmt, ANTARES, Swift, AGILE Team, 1M2H Team, Dark Energy Camera GW-EM, DES, DLT40, GRAWITA, Fermi-LAT, ATCA, ASKAP, Las Cumbres Observatory Group, OzGrav, DWF (Deeper Wider Faster Program), AST3, CAASTRO, VINROUGE, MASTER, J-GEM, GROWTH, JAGWAR, CaltechNRAO, TTU-NRAO, NuSTAR, Pan-STARRS, MAXI Team, TZAC Consortium, KU, Nordic Optical Telescope, ePESSTO, GROND, Texas Tech University, SALT Group, TOROS, BOOTES, MWA, CALET, IKI-GW Follow-up, H.E.S.S., LOFAR, LWA, HAWC, Pierre Auger, ALMA, Euro VLBI Team, Pi of Sky, Chandra Team at McGill University, DFN, ATLAS Telescopes, High Time Resolution Universe Survey, RIMAS, RATIR, SKA South Africa/MeerKAT",
    title = "{Multi-messenger Observations of a Binary Neutron Star Merger}",
    eprint = "1710.05833",
    archivePrefix = "arXiv",
    primaryClass = "astro-ph.HE",
    reportNumber = "LIGO-P1700294, VIR-0802A-17, FERMILAB-PUB-17-478-A-AE-CD",
    doi = "10.3847/2041-8213/aa91c9",
    journal = "Astrophys. J. Lett.",
    volume = "848",
    number = "2",
    pages = "L12",
    year = "2017"
}

@article{LIGOScientific:2017zic,
    author = "Abbott, B. P. and others",
    collaboration = "LIGO Scientific, Virgo, Fermi-GBM, INTEGRAL",
    title = "{Gravitational Waves and Gamma-rays from a Binary Neutron Star Merger: GW170817 and GRB 170817A}",
    eprint = "1710.05834",
    archivePrefix = "arXiv",
    primaryClass = "astro-ph.HE",
    reportNumber = "LIGO-P1700308",
    doi = "10.3847/2041-8213/aa920c",
    journal = "Astrophys. J. Lett.",
    volume = "848",
    number = "2",
    pages = "L13",
    year = "2017"
}

@article{Benhar:1998au,
    author = "Benhar, Omar and Berti, Emanuele and Ferrari, Valeria",
    editor = "Ferrari, V. and Miller, J. C. and Rezzolla, L.",
    title = "{The Imprint of the equation of state on the axial w modes of oscillating neutron stars}",
    eprint = "gr-qc/9901037",
    archivePrefix = "arXiv",
    doi = "10.1046/j.1365-8711.1999.02983.x",
    journal = "Mon. Not. Roy. Astron. Soc.",
    volume = "310",
    pages = "797--803",
    year = "1999"
}

@article{Miller:2019cac,
    author = "Miller, M. C. and others",
    title = "{PSR J0030+0451 Mass and Radius from $NICER$ Data and Implications for the Properties of Neutron Star Matter}",
    eprint = "1912.05705",
    archivePrefix = "arXiv",
    primaryClass = "astro-ph.HE",
    doi = "10.3847/2041-8213/ab50c5",
    journal = "Astrophys. J. Lett.",
    volume = "887",
    number = "1",
    pages = "L24",
    year = "2019"
}

@article{Bogdanov:2019ixe,
    author = "Bogdanov, Slavko and others",
    title = "{Constraining the Neutron Star Mass{\textendash}Radius Relation and Dense Matter Equation of State with $NICER$. I. The Millisecond Pulsar X-Ray Data Set}",
    eprint = "1912.05706",
    archivePrefix = "arXiv",
    primaryClass = "astro-ph.HE",
    doi = "10.3847/2041-8213/ab53eb",
    journal = "Astrophys. J. Lett.",
    volume = "887",
    number = "1",
    pages = "L25",
    year = "2019"
}

@article{Bogdanov:2019qjb,
    author = "Bogdanov, Slavko and others",
    title = "{Constraining the Neutron Star Mass{\textendash}Radius Relation and Dense Matter Equation of State with $NICER$. II. Emission from Hot Spots on a Rapidly Rotating Neutron Star}",
    eprint = "1912.05707",
    archivePrefix = "arXiv",
    primaryClass = "astro-ph.HE",
    doi = "10.3847/2041-8213/ab5968",
    journal = "Astrophys. J. Lett.",
    volume = "887",
    number = "1",
    pages = "L26",
    year = "2019"
}

@article{Bogdanov:2006zd,
    author = "Bogdanov, Slavko and Rybicki, George B. and Grindlay, Jonathan E.",
    title = "{Constraints on Neutron Star Properties from X-ray Observations of Millisecond Pulsars}",
    eprint = "astro-ph/0612791",
    archivePrefix = "arXiv",
    doi = "10.1086/520793",
    journal = "Astrophys. J.",
    volume = "670",
    pages = "668",
    year = "2007"
}

@article{Chatziioannou:2020pqz,
    author = "Chatziioannou, Katerina",
    title = "{Neutron star tidal deformability and equation of state constraints}",
    eprint = "2006.03168",
    archivePrefix = "arXiv",
    primaryClass = "gr-qc",
    doi = "10.1007/s10714-020-02754-3",
    journal = "Gen. Rel. Grav.",
    volume = "52",
    number = "11",
    pages = "109",
    year = "2020"
}

@article{Riley:2019yda,
    author = "Riley, Thomas E. and others",
    title = "{A $NICER$ View of PSR J0030+0451: Millisecond Pulsar Parameter Estimation}",
    eprint = "1912.05702",
    archivePrefix = "arXiv",
    primaryClass = "astro-ph.HE",
    doi = "10.3847/2041-8213/ab481c",
    journal = "Astrophys. J. Lett.",
    volume = "887",
    number = "1",
    pages = "L21",
    year = "2019"
}

@article{Miller:2021qha,
    author = "Miller, M. C. and others",
    title = "{The Radius of PSR J0740+6620 from NICER and XMM-Newton Data}",
    eprint = "2105.06979",
    archivePrefix = "arXiv",
    primaryClass = "astro-ph.HE",
    doi = "10.3847/2041-8213/ac089b",
    journal = "Astrophys. J. Lett.",
    volume = "918",
    number = "2",
    pages = "L28",
    year = "2021"
}

@article{Riley:2021pdl,
    author = "Riley, Thomas E. and others",
    title = "{A NICER View of the Massive Pulsar PSR J0740+6620 Informed by Radio Timing and XMM-Newton Spectroscopy}",
    eprint = "2105.06980",
    archivePrefix = "arXiv",
    primaryClass = "astro-ph.HE",
    doi = "10.3847/2041-8213/ac0a81",
    journal = "Astrophys. J. Lett.",
    volume = "918",
    number = "2",
    pages = "L27",
    year = "2021"
}

@article{Akmal:1998cf,
    author = "Akmal, A. and Pandharipande, V. R. and Ravenhall, D. G.",
    title = "{The Equation of state of nucleon matter and neutron star structure}",
    eprint = "nucl-th/9804027",
    archivePrefix = "arXiv",
    doi = "10.1103/PhysRevC.58.1804",
    journal = "Phys. Rev. C",
    volume = "58",
    pages = "1804--1828",
    year = "1998"
}

@article{LIGOScientific:2018cki,
    author = "Abbott, B. P. and others",
    collaboration = "LIGO Scientific, Virgo",
    title = "{GW170817: Measurements of neutron star radii and equation of state}",
    eprint = "1805.11581",
    archivePrefix = "arXiv",
    primaryClass = "gr-qc",
    reportNumber = "LIGO-P1800115",
    doi = "10.1103/PhysRevLett.121.161101",
    journal = "Phys. Rev. Lett.",
    volume = "121",
    number = "16",
    pages = "161101",
    year = "2018"
}

@article{Diaz-Guerra:2024gff,
    author = "D{\'\i}az-Guerra, David and Albertus, Conrado and Char, Prasanta and P{\'e}rez-Garc{\'\i}a, M. {\'A}ngeles",
    title = "{Gauge-invariant perturbations of relativistic nonperfect fluids in spherical spacetime}",
    eprint = "2410.18081",
    archivePrefix = "arXiv",
    primaryClass = "gr-qc",
    doi = "10.1103/PhysRevD.110.124012",
    journal = "Phys. Rev. D",
    volume = "110",
    number = "12",
    pages = "124012",
    year = "2024"
}

@article{Clarisse:2025lli,
    author = "Clarisse, Nicolas and Pinho, Eduardo O. and Patel, Teerthal and Bemfica, Fabio S. and Hippert, Mauricio and Noronha, Jorge",
    title = "{Flux-conservative Bemfica-Disconzi-Noronha-Kovtun hydrodynamics and shock regularization}",
    eprint = "2510.16603",
    archivePrefix = "arXiv",
    primaryClass = "gr-qc",
    doi = "10.1103/f8y1-3yck",
    journal = "Phys. Rev. D",
    volume = "113",
    number = "2",
    pages = "024051",
    year = "2026"
}

@article{Chomali-Castro:2026qww,
    author = "Chomal{\'\i}-Castro, Vicente and Clarisse, Nick and Mullins, Nicki and Noronha, Jorge",
    title = "{Solving BDNK diffusion using physics-informed neural networks}",
    eprint = "2602.16117",
    archivePrefix = "arXiv",
    primaryClass = "nucl-th",
    month = "2",
    year = "2026"
}

@article{Bantilan:2022ech,
    author = "Bantilan, Hans and Bea, Yago and Figueras, Pau",
    title = "{Evolutions in first-order viscous hydrodynamics}",
    eprint = "2201.13359",
    archivePrefix = "arXiv",
    primaryClass = "hep-th",
    reportNumber = "HIP-2022-1/TH",
    doi = "10.1007/JHEP08(2022)298",
    journal = "JHEP",
    volume = "08",
    pages = "298",
    year = "2022"
}

@article{Keeble:2025bkc,
    author = "Keeble, Lennox S. and Pretorius, Frans",
    title = "{First-order viscous relativistic hydrodynamics on the two-sphere}",
    eprint = "2508.20998",
    archivePrefix = "arXiv",
    primaryClass = "gr-qc",
    doi = "10.1103/d4wd-zj7w",
    journal = "Phys. Rev. D",
    volume = "112",
    number = "12",
    pages = "124034",
    year = "2025"
}

@article{Israel:1979wp,
    author = "Israel, W. and Stewart, J. M.",
    title = "{Transient relativistic thermodynamics and kinetic theory}",
    doi = "10.1016/0003-4916(79)90130-1",
    journal = "Annals Phys.",
    volume = "118",
    pages = "341--372",
    year = "1979"
}

@Inbook{vonNeumann,
author="von Neumann, J.
and Wigner, E. P.",
editor="Wightman, Arthur S.",
title="{\"U}ber das Verhalten von Eigenwerten bei adiabatischen Prozessen",
bookTitle="The Collected Works of Eugene Paul Wigner: Part A: The Scientific Papers",
year="1993",
publisher="Springer Berlin Heidelberg",
address="Berlin, Heidelberg",
pages="294--297",
isbn="978-3-662-02781-3",
doi="10.1007/978-3-662-02781-3_20",
url="https://doi.org/10.1007/978-3-662-02781-3_20"
}

@article{Christensen:2003,
  title={Lecture notes on stellar oscillations},
  author={Christensen-Dalsgaard, J{\o}rgen},
  year={2003},
  publisher={Institut for Fysik og Astronomi, Aarhus Universitet}
}

@article{Oshita:2025ibu,
    author = "Oshita, Naritaka and Berti, Emanuele and Cardoso, Vitor",
    title = "{Unstable Chords and Destructive Resonant Excitation of Black Hole Quasinormal Modes}",
    eprint = "2503.21276",
    archivePrefix = "arXiv",
    primaryClass = "gr-qc",
    reportNumber = "YITP-25-44, RIKEN-iTHEMS-Report-25",
    doi = "10.1103/ht2n-vvvh",
    journal = "Phys. Rev. Lett.",
    volume = "135",
    number = "3",
    pages = "031401",
    year = "2025"
}

@article{PanossoMacedo:2025xnf,
    author = "Panosso Macedo, Rodrigo and Katagiri, Takuya and Kubota, Kei-ichiro and Motohashi, Hayato",
    title = "{Exceptional Points and Resonance in Black Hole Ringdown}",
    eprint = "2512.02110",
    archivePrefix = "arXiv",
    primaryClass = "gr-qc",
    month = "12",
    year = "2025"
}

@article{Oshita:2021iyn,
    author = "Oshita, Naritaka",
    title = "{Ease of excitation of black hole ringing: Quantifying the importance of overtones by the excitation factors}",
    eprint = "2109.09757",
    archivePrefix = "arXiv",
    primaryClass = "gr-qc",
    reportNumber = "RIKEN-iTHEMS-Report-21",
    doi = "10.1103/PhysRevD.104.124032",
    journal = "Phys. Rev. D",
    volume = "104",
    number = "12",
    pages = "124032",
    year = "2021"
}

@article{Motohashi:2024fwt,
    author = "Motohashi, Hayato",
    title = "{Resonant Excitation of Quasinormal Modes of Black Holes}",
    eprint = "2407.15191",
    archivePrefix = "arXiv",
    primaryClass = "gr-qc",
    doi = "10.1103/PhysRevLett.134.141401",
    journal = "Phys. Rev. Lett.",
    volume = "134",
    number = "14",
    pages = "141401",
    year = "2025"
}

@article{Onozawa:1996ux,
    author = "Onozawa, Hisashi",
    title = "{A Detailed study of quasinormal frequencies of the Kerr black hole}",
    eprint = "gr-qc/9610048",
    archivePrefix = "arXiv",
    reportNumber = "TIT-HEP-344, COSMO-78",
    doi = "10.1103/PhysRevD.55.3593",
    journal = "Phys. Rev. D",
    volume = "55",
    pages = "3593--3602",
    year = "1997"
}

@article{Cardoso:2019rvt,
    author = "Cardoso, Vitor and Pani, Paolo",
    title = "{Testing the nature of dark compact objects: a status report}",
    eprint = "1904.05363",
    archivePrefix = "arXiv",
    primaryClass = "gr-qc",
    doi = "10.1007/s41114-019-0020-4",
    journal = "Living Rev. Rel.",
    volume = "22",
    number = "1",
    pages = "4",
    year = "2019"
}

@article{LIGOScientific:2026wpt,
    author = "Abac, A. G. and others",
    collaboration = "LIGO Scientific, VIRGO, KAGRA",
    title = "{GWTC-4.0: Tests of General Relativity. III. Tests of the Remnants}",
    eprint = "2603.19021",
    archivePrefix = "arXiv",
    primaryClass = "gr-qc",
    reportNumber = "LIGO-P2500067",
    month = "3",
    year = "2026"
}

@article{Leaver:1985ax,
    author = "Leaver, E. W.",
    title = "{An Analytic representation for the quasi normal modes of Kerr black holes}",
    doi = "10.1098/rspa.1985.0119",
    journal = "Proc. Roy. Soc. Lond. A",
    volume = "402",
    pages = "285--298",
    year = "1985"
}

@article{Leaver:1986vnb,
    author = "Leaver, E. W.",
    title = "{Solutions to a generalized spheroidal wave equation: Teukolsky{\textquoteright}s equations in general relativity, and the two-center problem in molecular quantum mechanics}",
    doi = "10.1063/1.527130",
    journal = "J. Math. Phys.",
    volume = "27",
    number = "5",
    pages = "1238",
    year = "1986"
}

@article{Leaver:1990zz,
    author = "Leaver, Edward W.",
    title = "{Quasinormal modes of Reissner-Nordstrom black holes}",
    doi = "10.1103/PhysRevD.41.2986",
    journal = "Phys. Rev. D",
    volume = "41",
    pages = "2986--2997",
    year = "1990"
}

@article{Gautschi:1967,
  title={Computational aspects of three-term recurrence relations},
  author={Gautschi, Walter},
  journal={SIAM review},
  volume={9},
  number={1},
  pages={24--82},
  year={1967},
  publisher={SIAM}
}

@article{Siemonsen:2025fne,
    author = "Siemonsen, Nils",
    title = "{Weakly turbulent saturation of the nonlinear scalar ergoregion instability}",
    eprint = "2510.07467",
    archivePrefix = "arXiv",
    primaryClass = "gr-qc",
    month = "10",
    year = "2025"
}

@article{Siemonsen:2025wib,
    author = "Siemonsen, Nils",
    title = "{Ergoregion instability in bosonic stars: Scalar mode structure, universality, and weakly nonlinear effects}",
    eprint = "2510.07468",
    archivePrefix = "arXiv",
    primaryClass = "gr-qc",
    doi = "10.1103/sbtt-jnjg",
    journal = "Phys. Rev. D",
    volume = "112",
    number = "12",
    pages = "124031",
    year = "2025"
}

@article{Buchdahl:1959zz,
    author = "Buchdahl, Hans A.",
    title = "{General Relativistic Fluid Spheres}",
    doi = "10.1103/PhysRev.116.1027",
    journal = "Phys. Rev.",
    volume = "116",
    pages = "1027",
    year = "1959"
}

@article{Benomio:2018ivy,
    author = "Benomio, Gabriele",
    title = "{The stable trapping phenomenon for black strings and black rings and its obstructions on the decay of linear waves}",
    eprint = "1809.07795",
    archivePrefix = "arXiv",
    primaryClass = "gr-qc",
    doi = "10.2140/apde.2021.14.2427",
    journal = "Anal. Part. Diff. Eq.",
    volume = "14",
    number = "8",
    pages = "2427--2496",
    year = "2021"
}

@article{Evstafyeva:2025mvx,
    author = "Evstafyeva, Tamara and Siemonsen, Nils and East, William E.",
    title = "{Assessing the stability of ultracompact spinning boson stars with nonlinear evolutions}",
    eprint = "2508.11527",
    archivePrefix = "arXiv",
    primaryClass = "gr-qc",
    doi = "10.1103/2vz3-s39r",
    journal = "Phys. Rev. D",
    volume = "113",
    number = "4",
    pages = "044024",
    year = "2026"
}

@article{Siemonsen:2024snb,
    author = "Siemonsen, Nils",
    title = "{Nonlinear Treatment of a Black Hole Mimicker Ringdown}",
    eprint = "2404.14536",
    archivePrefix = "arXiv",
    primaryClass = "gr-qc",
    doi = "10.1103/PhysRevLett.133.031401",
    journal = "Phys. Rev. Lett.",
    volume = "133",
    number = "3",
    pages = "031401",
    year = "2024"
}

@article{Cunha:2022gde,
    author = "Cunha, Pedro V. P. and Herdeiro, Carlos and Radu, Eugen and Sanchis-Gual, Nicolas",
    title = "{Exotic Compact Objects and the Fate of the Light-Ring Instability}",
    eprint = "2207.13713",
    archivePrefix = "arXiv",
    primaryClass = "gr-qc",
    doi = "10.1103/PhysRevLett.130.061401",
    journal = "Phys. Rev. Lett.",
    volume = "130",
    number = "6",
    pages = "061401",
    year = "2023"
}

@article{Dias:2022oqm,
    author = "Dias, Oscar J. C. and Godazgar, Mahdi and Santos, Jorge E.",
    title = "{Eigenvalue repulsions and quasinormal mode spectra of Kerr-Newman: an extended study}",
    eprint = "2205.13072",
    archivePrefix = "arXiv",
    primaryClass = "gr-qc",
    doi = "10.1007/JHEP07(2022)076",
    journal = "JHEP",
    volume = "07",
    pages = "076",
    year = "2022"
}

@article{Berti:2025hly,
    author = "Abedi, Jahed and others",
    editor = "Berti, Emanuele and Cardoso, Vitor and Carullo, Gregorio",
    title = "{Black hole spectroscopy: from theory to experiment}",
    eprint = "2505.23895",
    archivePrefix = "arXiv",
    primaryClass = "gr-qc",
    month = "5",
    year = "2025"
}

\end{document}